\documentclass[12pt,a4paper]{article}
\usepackage{graphics,graphicx}
\usepackage{amssymb,epsfig,amsmath,euscript,array}
\usepackage{cite}

\makeatletter
\@addtoreset{equation}{section}
\makeatother
\renewcommand{\theequation}{\thesection.\arabic{equation}}

\newcommand{\startappendix}{
\setcounter{section}{0}
\renewcommand{\thesection}{\Alph{section}}
\renewcommand{\theequation}{\Alph{section}.\arabic{equation}}}

\newcounter{multieqs}

\newenvironment{pretty}{}{}

\newcommand{\be}{\begin{equation}}
\newcommand{\ee}{\end{equation}}

\newcommand{\bm}[1]{\mbox{\boldmath $#1$}}

\newcommand{\kslash}{k \!\!\! / }

\newcommand{\lslash}{l \!\! / }
\newcommand{\Pslash}{P \!\!\!\! / }

\newcommand{\islash}{i \!\!\! / }
\newcommand{\jslash}{j \!\!\! / }
\newcommand{\aslash}{a \!\!\! / }
\newcommand{\bslash}{{b \hspace{-6pt} \slash} }

\newcommand{\onslash}{1 \!\!\! / }
\newcommand{\twslash}{2 \!\!\!/ }
\newcommand{\thslash}{3 \!\!\!/ }
\newcommand{\foslash}{4 \!\!\! / }
\newcommand{\fislash}{5 \!\!\! / }

\newcommand{\mslash}{m \!\!\! / }

\def\bd{\begin{document}}
\def\ed{\end{document}}
\def\nn{\nonumber}
\def\bea{\begin{eqnarray}}
\def\eea{\end{eqnarray}}
\def\eps{\epsilon}

\def\ab{(ijab)}
\def\ba{(ijba)}
\def\ijab{{\tr}_{+}(\islash\, \jslash\, \aslash \, \bslash)}
\def\ijba{{\tr}_{+}(\islash\, \jslash\, \bslash \, \aslash)}
\def\ijaP{{\tr}_{+}(\islash\, \jslash\, \aslash \, \Pslash)}
\def\ijPLa{{\tr}_{+}(\islash\, \jslash\, \Pslash_L \, \aslash)}
\def\ijaPL{{\tr}_{+}(\islash\, \jslash\, \aslash \, \Pslash_L)}
\def\ijPLza{{\tr}_{+}(\islash\, \jslash\, \Pslash_{L;z} \, \aslash)}
\def\ijaPLz{{\tr}_{+}(\islash\, \jslash\, \aslash \, \Pslash_{L;z})}
\def\ijPa{{\tr}_{+}(\islash\, \jslash\, \Pslash \, \aslash)}
\def\iaPb{{\tr}_{+}(\islash\, \aslash\, \Pslash \, \bslash)}
\def\ibPa{{\tr}_{+}(\islash\, \bslash\, \Pslash \, \aslash)}
\def\ijPmu{{\tr}_{+}(\islash\, \jslash\, \Pslash \, \mu)}
\def\ibmuP{{\tr}_{+}(\islash\, \bslash\, \mu \, \Pslash)}
\def\ibmua{{\tr}_{+}(\islash\, \bslash\, \mu \, \aslash)}
\def\iamub{{\tr}_{+}(\islash\, \aslash\, \mu \, \bslash)}
\def\jaPb{{\tr}_{+}(\jslash\, \aslash\, \Pslash \, \bslash)}
\def\ijmuP{{\tr}_{+}(\islash\, \jslash\, \mu \, \Pslash)}
\def\ijmum{{\tr}_{+}(\islash\, \jslash\, \mu \, \mslash)}
\def\ijmmu{{\tr}_{+}(\islash\, \jslash\, \mslash \, \mu)}
\def\ijmP{{\tr}_{+}(\islash\, \jslash\, \mslash \, \Pslash)}
\def\iabP{{\tr}_{+}(\islash\, \aslash\, \bslash \, \Pslash)}
\def\ijbP{{\tr}_{+}(\islash\, \jslash\, \bslash \, \Pslash)}
\def\jbPa{{\tr}_{+}(\jslash\, \bslash\, \Pslash \, \aslash)}
\def\ijPb{{\tr}_{+}(\islash\, \jslash\, \Pslash \, \bslash)}
\def\jbmua{{\tr}_{+}(\jslash\, \bslash\, \mu \, \aslash)}

\def\loablt{ {\tr}_{+}(\lslash_1\, \aslash \, \bslash\, \lslash_2)}

\def\ijlolt{{\tr}_{+}(\islash\, \jslash\, \lslash_1 \, \lslash_2)}
\def\ijltlo{{\tr}_{+}(\islash\, \jslash\, \lslash_2 \, \lslash_1)}
\def\ibloa{{\tr}_{+}(\islash\, \bslash\, \lslash_1 \, \aslash)}
\def\jaltb{{\tr}_{+}(\jslash\, \aslash\, \lslash_2 \, \bslash)}
\def\ialtb{{\tr}_{+}(\islash\, \aslash\, \lslash_2 \, \bslash)}
\def\bltloa{{\tr}_{+}(\bslash\, \lslash_2\, \lslash_1 \, \aslash)}
\def\jbloa{{\tr}_{+}(\jslash\, \bslash\, \lslash_1 \, \aslash)}
\def\ibPb{{\tr}_{+}(\islash\, \bslash\, \Pslash \, \bslash)}
\def\ijltb{{\tr}_{+}(\islash\, \jslash\, \lslash_2 \, \bslash)}

\def\ijloa{{\tr}_{+}(\islash\, \jslash\,  \lslash_1 \, \aslash)}
\def\ijblt{{\tr}_{+}(\islash\, \jslash\,  \bslash \, \lslash_2)}

\def\jakb{{\tr}_{+}(\jslash\, \aslash\, \kslash \, \bslash)}
\def\iakb{{\tr}_{+}(\islash\, \aslash\, \kslash \, \bslash)}

\def\tofo{{\tr}_{+}(\onslash\, \thslash\, \twslash \, \foslash)}
\def\foto{{\tr}_{+}(\onslash\, \thslash\, \foslash \, \twslash)}
\def\tofi{{\tr}_{+}(\onslash\, \thslash\, \twslash \, \fislash)}
\def\fito{{\tr}_{+}(\onslash\, \thslash\, \fislash \, \twslash)}

\newcommand{\note}[2]{{\footnotesize [#1}---{\footnotesize \sc  #2]}}

\def\lrangle#1#2{\langle #1\,#2\rangle}

\def\Li{{$\rm Li}_2$}

\let\bm=\bibitem
\let\la=\label

\def\npb#1#2#3{Nucl. Phys. {\bf{B#1}} #3 (#2)}
\def\plb#1#2#3{Phys. Lett. {\bf{#1B}} #3 (#2)}
\def\prl#1#2#3{Phys. Rev. Lett. {\bf{#1}} #3 (#2)}
\def\prd#1#2#3{Phys. Rev. {D \bf{#1}} #3 (#2)}
\def\cmp#1#2#3{Comm. Math. Phys. {\bf{#1}} #3 (#2)}
\def\cqg#1#2#3{Class. Quantum Grav. {\bf{#1}} #3 (#2)}
\def\nppsa#1#2#3{Nucl. Phys. B (Proc. Suppl.) {\bf{#1A}}#3 (#2)}
\def\ap#1#2#3{Ann. of Phys. {\bf{#1}} #3 (#2)}
\def\ijmp#1#2#3{Int. J. Mod. Phys. {\bf{A#1}} #3 (#2)}
\def\rmp#1#2#3{Rev. Mod. Phys. {\bf{#1}} #3 (#2)}
\def\mpla#1#2#3{Mod. Phys. Lett. {\bf A#1} #3 (#2)}
\def\jhep#1#2#3{J. High Energy Phys. {\bf #1} #3 (#2)}
\def\atmp#1#2#3{Adv. Theor. Math. Phys. {\bf #1} #3 (#2)}
\newcommand{\EQ}[1]{\begin{equation} #1 \end{equation}}
\newcommand{\AL}[1]{\begin{subequations}\begin{align} #1 \end{align}\end{subequations}}
\newcommand{\SP}[1]{\begin{equation}\begin{split} #1 \end{split}\end{equation}}
\newcommand{\ALAT}[2]{\begin{subequations}\begin{alignat}{#1} #2 \end{alignat}
                        \end{subequations}}
\def\beqa{\begin{eqnarray}}
\def\eeqa{\end{eqnarray}}
\def\beq{\begin{equation}}
\def\eeq{\end{equation}}
\def\sst{\scriptscriptstyle}
\def\thetabar{\bar\theta}
\def\Tr{{\rm Tr}}
\def\one{\mbox{1 \kern-.59em {\rm l}}}
 \def\Nh{\hat{N}}

\def\a{\alpha}      \def\da{{\dot\alpha}}
\def\b{\beta}       \def\db{{\dot\beta}}
\def\g{\gamma}  \def\G{\Gamma}  \def\cdt{\dot\gamma}
\def\d{\delta}  \def\D{\Delta}  \def\ddt{\dot\delta}
\def\e{\epsilon}        \def\vare{\varepsilon}
\def\f{\phi}    \def\F{\Phi}    \def\vvf{\f}
\def\h{\eta}
\def\k{\kappa}
\def\l{\lambda} \def\L{\Lambda}
\def\m{\mu} \def\n{\nu}
\def\o{\omega}
\def\p{\pi} \def\P{\Pi}
\def\r{\rho}
\def\s{\sigma}  \def\S{\Sigma}
\def\t{\tau}
\def\th{\theta} \def\Th{\Theta} \def\vth{\vartheta}
\def\X{\Xeta}
\def\z{\zeta}

\def\cA{{\cal A}} \def\cB{{\cal B}} \def\cC{{\cal C}}
\def\cD{{\cal D}} \def\cE{{\cal E}} \def\cF{{\cal F}}
\def\cG{{\cal G}} \def\cH{{\cal H}} \def\cI{{\cal I}}
\def\cJ{{\cal J}} \def\cK{{\cal K}} \def\cL{{\cal L}}
\def\cM{{\cal M}} \def\cN{{\cal N}} \def\cO{{\cal O}}
\def\cP{{\cal P}} \def\cQ{{\cal Q}} \def\cR{{\cal R}}
\def\cS{{\cal S}} \def\cT{{\cal T}} \def\cU{{\cal U}}
\def\cV{{\cal V}} \def\cW{{\cal W}} \def\cX{{\cal X}}
\def\cY{{\cal Y}} \def\cZ{{\cal Z}}

\def\ua{\underline{\alpha}}
\def\ub{\underline{\phantom{\alpha}}\!\!\!\beta}
\def\uc{\underline{\phantom{\alpha}}\!\!\!\gamma}
\def\um{\underline{\mu}}
\def\ud{\underline\delta}
\def\ue{\underline\epsilon}
\def\una{\underline a}\def\unA{\underline A}
\def\unb{\underline b}\def\unB{\underline B}
\def\unc{\underline c}\def\unC{\underline C}
\def\und{\underline d}\def\unD{\underline D}
\def\une{\underline e}\def\unE{\underline E}
\def\unf{\underline{\phantom{e}}\!\!\!\! f}\def\unF{\underline F}
\def\unm{\underline m}\def\unM{\underline M}
\def\unn{\underline n}\def\unN{\underline N}
\def\unp{\underline{\phantom{a}}\!\!\! p}\def\unP{\underline P}
\def\unq{\underline{\phantom{a}}\!\!\! q}
\def\unQ{\underline{\phantom{A}}\!\!\!\! Q}
\def\unH{\underline{H}}

\def\As {{A \hspace{-6.4pt} \slash}\;}
\def\bs {{b \hspace{-6.4pt} \slash}\;}
\def\Ds {{D \hspace{-6.4pt} \slash}\;}
\def\ds {{\del \hspace{-6.4pt} \slash}\;}
\def\ss {{\s \hspace{-6.4pt} \slash}\;}
\def\ks {{ k \hspace{-6.4pt} \slash}\;}
\def\ps {{p \hspace{-6.4pt} \slash}\;}
\def\pas {{{p_1} \hspace{-6.4pt} \slash}\;}
\def\pbs {{{p_2} \hspace{-6.4pt} \slash}\;}
\def\Ps {{P \hspace{-6.4pt} \slash}\;}
\def\Qs {{Q \hspace{-6.4pt} \slash}\;}

\def\Fh{\hat{F}}
\def\Vh{\hat{V}}
\def\Xh{\hat{X}}
\def\ah{\hat{a}}
\def\xh{\hat{x}}
\def\yh{\hat{y}}
\def\ph{\hat{p}}
\def\xih{\hat{\xi}}
\def\psit{\tilde{\psi}}
\def\Psit{\tilde{\Psi}}
\def\tht{\tilde{\th}}
\def\lt{\tilde{\lambda}}
\def\llt{\tilde{l}}
\def\At{\tilde{A}}
\def\Qt{\tilde{Q}}
\def\Rt{\tilde{R}}
\def\Nt{\tilde{N}}

\def\at{\tilde{a}}
\def\st{\tilde{s}}
\def\ft{\tilde{f}}
\def\pt{\tilde{p}}
\def\qt{\tilde{q}}
\def\vt{\tilde{v}}
\def\nt{\tilde{n}}

\def\delb{\bar{\partial}}
\def\bz{\bar{z}}
\def\bD{\bar{D}}
\def\bB{\bar{B}}

\def\bk{{\bf k}}
\def\bl{{\bf l}}
\def\bp{{\bf p}}
\def\bq{{\bf q}}
\def\br{{\bf r}}
\def\bx{{\bf x}}
\def\by{{\bf y}}
\def\bR{{\bf R}}
\def\bV{{\bf V}}

\def\d{\delta}\def\D{\Delta}\def\ddt{\dot\delta}
\def\pa{\partial} \def\del{\partial}
\def\xx{\times}
\def\uno{\mbox{1 \kern-.59em {\rm l}}}
\def\trp{^{\top}}
\def\inv{^{-1}}
\def\dag{{^{\dagger}}}
\def\pr{^{\prime}}
\def\lan{\langle}
\def\ran{\rangle}
\def\rar{\rightarrow}
\def\lar{\leftarrow}
\def\lrar{\leftrightarrow}
\newcommand{\0}{\,\!}      
\def\one{1\!\!1\,\,}
\def\im{\imath}
\def\jm{\jmath}
\newcommand{\tr}{\mbox{tr}}
\newcommand{\slsh}[1]{/ \!\!\!\! #1}
\def\vac{|0\rangle}
\def\lvac{\langle 0|}
\def\hlf{\frac{1}{2}}
\def\ove#1{\frac{1}{#1}}
\def\Box{\square}
\def\ZZ{\mathbb{Z}}
\def\CC#1{({\bf #1})}
\def\bcomment#1{}
\def\bfhat#1{{\bf \hat{#1}}}
\def\VEV#1{\left\langle #1\right\rangle}
\newcommand{\ex}[1]{{\rm e}^{#1}} \def\ii{{\rm i}}
\def\rr{{\rm r}} \def\rs{{\rm s}}\def\rv{{\rm v}}
\def\ri{{\rm i}}\def\rj{{\rm j}}
\newcommand{\lrbrk}[1]{\left(#1\right)}
\newcommand{\sfrac}[2]{{\textstyle\frac{#1}{#2}}}

\def\Li{{\rm Li}_2}

\def\Li2{{\rm Li}_2}

\def\intD{\int\!{d^DL \over (2\pi)^D}}

\font\mybb=msbm10 at 12pt
\def\bb#1{\hbox{\mybb#1}}

\font\myBB=msbm10 at 18pt
\def\BB#1{\hbox{\myBB#1}}

\begin{document}

\begin{flushright}
QMUL-PH-07-10
\end{flushright}

\vspace{20pt}

\begin{center}

{\Large \bf  One-loop  $\cN\!=\!8$  Supergravity  Amplitudes}\\
\vspace{0.3cm}
{\Large \bf from  MHV Diagrams }
\vspace{33pt}

{\bf {\mbox{Adele Nasti and Gabriele  Travaglini}}}%
\begin{pretty}
\footnote{{\sffamily \{\tt a.nasti, g.travaglini\}@qmul.ac.uk }}
\end{pretty}

{\em Centre for Research in String Theory \\ Department of Physics\\
Queen Mary, University of
London\\
Mile End Road, London, E1 4NS\\
United Kingdom}
\vspace{40pt}

{\bf Abstract}

\end{center}

\noindent
We discuss the calculation of one-loop amplitudes in $\cN\!=\!8$ supergravity
using MHV diagrams. In contrast to MHV amplitudes of gluons in Yang-Mills,
tree-level MHV amplitudes of gravitons are not holomorphic in the spinor variables.
In order to extend  these amplitudes off shell, and use them as vertices to build  loops,
we introduce certain shifts for the spinor variables associated to the loop momenta.
Using this off-shell prescription, we rederive the four-point
MHV amplitude of gravitons at one loop, in complete agreement with known results.
We also discuss the extension to the case of one-loop MHV amplitudes with
an arbitrary number of  gravitons.

\vspace{0.5cm}

\setcounter{page}{0}
\thispagestyle{empty}
\newpage


\setcounter{footnote}{0}

\section{Introduction}

Over the past years, several new techniques in perturbative quantum field theory
have emerged, following Witten's proposal that
weakly-coupled Yang-Mills theory can be equivalently described by a
twistor string theory \cite{witten} (see \cite{Cachazo:2005ga} for a review).
The first twistor-inspired realisation of a diagrammatic method alternative
to Feynman diagrams is the MHV diagram method introduced by Cachazo, Svr\v{c}ek and Witten (CSW)
in \cite{csw}. In that paper, it was proposed that MHV scattering amplitudes of gluons
appropriately continued off the mass shell, can be used as vertices, to be joined
with scalar propagators, in a novel perturbative expansion of Yang-Mills theory.
The  proposal of CSW,  originally applied to amplitudes at tree level,
was strongly supported  by the study of the multi-particle singularities of
the amplitudes, which are neatly reproduced by a calculation based on MHV diagrams.
Shortly after, several old and new amplitudes at tree level were computed
in \cite{v1,z1,z2,v2}, also with fermions and scalars on the external legs.

A key  ingredient of the CSW approach is the introduction of an off-shell
continuation of the Parke-Taylor formula for the MHV amplitude of gluons,
which  is necessary in order to lift the amplitude to a full-fledged vertex.
This off-shell continuation, which we will discuss in the following sections,
is based on a decomposition of momenta identical to that used in
lightcone quantisation of Yang-Mills, where a generic momentum $L$ is written as
$L = l + z \eta$. Here $\eta_{\alpha \dot{\alpha}}:= \eta_\alpha \tilde{\eta}_{\dot{\alpha}}$
is  an arbitrary null vector
determining a lightlike direction, and  $l$ is also null.
This resemblance is not  accidental -- indeed,  it was shown
in \cite{Mansfield} (see also \cite{Gorsky:2005sf}) that a particular change of variables
in the lightcone Yang-Mills path integral leads to a  new action for the theory with  an infinite
number of MHV vertices.
Mansfield used   the holomorphicity of
the MHV amplitudes of gluons to argue that the new vertices are precisely given
by the Parke-Taylor formula continued off shell as proposed in \cite{csw}.
This was also checked explicitly for the four- and five-point vertices in \cite{Ettle:2006bw}.
The same off-shell prescription was also recently seen to emerge from   twistor
actions in an axial gauge in \cite{bms}. 

Although initially limited to Yang-Mills theory,   progress has also been
made in other theories, specifically in gravity.
This includes applications of the BCF recursion relation \cite{bcf,bcfw}
to amplitudes of gravitons, \cite{bbst3,cs,bmst2,Benincasa:2007qj}, and
(generalised) unitarity \cite{Bjerrum-Bohr:2005xx,zero}.%
\footnote{In particular, in \cite{zero}  (see also \cite{zerouno,Bjerrum-Bohr:2005xx})
the interesting hypothesis that  
one-loop $\cN=8$ supergravity amplitudes can be expanded in terms of scalar box functions only
was suggested. 
This hints at the  possibility that $\cN=8$ supergravity, similarly to 
$\cN=4$ super Yang-Mills,  is ultraviolet finite in four dimensions 
\cite{zero,Green:2006gt,Bern:2006kd,Green:2006yu,Bern:2007hh,chalmers}.}
 An important step was made in \cite{bdipr}, where MHV rules for tree-level gravity
amplitudes were formulated.%
\footnote{Earlier attempts at determining off-shell continuations of the
gravity MHV amplitudes  can be found in \cite{grav1,z2}.}
The strategy followed in that paper was to determine these MHV rules
as a special case of a BCF recursion relation, following the insight of \cite{ris}
for Yang-Mills theory.
For example, consider the calculation of a next-to-MHV   amplitude. By introducing 
shifts for the antiholomorphic spinors associated to the negative-helicity gluons, 
one obtains recursive diagrams immediately matching those of the CSW rules \cite{ris}.
Moreover, since gluon MHV amplitudes are holomorphic in the spinor variables,
these shifts are to all effects invisible in the gluon MHV vertex.
Finally, the spinor associated to the internal leg joining the two vertices
as dictated by the BCF recursion relation is nothing but that introduced in the
CSW prescription.
A similar picture emerged in gravity \cite{bdipr}, with the noticeable difference that
graviton MHV amplitudes depend explicitly upon antiholomorphic spinors, hence the precise form of
the shifts of \cite{ris} is very relevant. We note in passing that these shifts
break the reality condition $\bar{\lambda} = \pm \lambda^{\ast}$,
thereby leading naturally to a formulation
of (tree-level) MHV rules in complexified Minkowski space. The new tree-level
MHV rules of \cite{bdipr} were successfully used to derive explicit expressions 
for several amplitudes in General Relativity.

At the quantum level, the first applications of MHV rules were considered in
\cite{bst}, where the infinite sequence of one-loop MHV amplitude in $\cN=4$ super Yang-Mills 
was rederived using MHV diagrams (see \cite{BT06} for a review).
One of the main points of \cite{bst} is the derivation
of an expression for the loop integration measure, to be reviewed in section 2,
which made explicit the physical interpretation of the calculation as well as its
relation to the unitarity-based approach of Bern, Dixon, Dunbar and Kosower
\cite{bddk,bddkcoll}. This integration measure turned out to be the product of
a two-particle Lorentz-invariant phase space (LIPS) measure, and a dispersive measure.
In brief, one could summarise the essence of the method by saying that,
firstly,  the LIPS integration computes the discontinuity of the amplitude,
and then the dispersion integral reconstructs the full amplitude from its cuts.
In \cite{ftt}, it was shown using the local character of MHV vertices and the Feynman Tree Theorem 
\cite{F1,F2} 
that one-loop Yang-Mills amplitudes calculated using MHV diagrams 
are independent of the choice of the reference spinor  $\eta_{\alpha}$, and that, 
in the presence of supersymmetry, the correct collinear and soft singularities are reproduced, 
lending strong support to the  correctness of the method at one loop. 
Other applications of the method include the infinite sequence of MHV amplitudes in $\cN=1$ 
super Yang-Mills   \cite{quig,bbst1} and the cut-constructible part of the same amplitudes in 
pure Yang-Mills \cite{bbst2}, as well as the  recent calculations   \cite{Nigel1,Nigel2,Nigel3} of 
Higgs plus multi-gluon scattering amplitudes at one loop using the $\phi$-MHV rules introduced in   
\cite{LanceNigelValya} and further discussed in \cite{Badger:2004ty}.  
Amplitudes in non-supersymmetric Yang-Mills were also recently studied in 
\cite{pureYM,Ettle:2007qc,bstz}, where derivations of the finite all-minus and all-plus gluon amplitudes
were presented.

In this paper we will discuss the MHV diagram calculation  of the simplest one-loop amplitudes
in gravity, namely the MHV amplitudes of gravitons in maximally supersymmetric
$\cN=8$ supergravity. The four-point amplitude, which we will reproduce in detail,
was first obtained from the $\alpha^{\prime} \to 0$ limit of
a string theory calculation in \cite{Green:1982sw}, and then rederived in
\cite{dn} with the string-based method of  \cite{bk}, and also  using  unitarity. 
The infinite sequence of MHV amplitudes was later
obtained  in \cite{Bern:1998sv}.%
\footnote{ See \cite{Bern:2002kj} for a nice review on gravity amplitudes and their properties.} 
By construction, two-particle cuts and generalised cuts 
of a generic one-loop gravity amplitude obtained using an MHV diagram based 
approach automatically agree with those of the correct amplitude, 
in complete similarity  to  the Yang-Mills case (see the discussion in section 4 of \cite{ftt}). 
As in Yang-Mills, the crux of the problem will be
determining the off-shell continuation of the spinors associated to the loop legs,
which will affect the rational terms in the amplitude; this off-shell continuation should 
be such that  the final result is independent
of the particular choice of the reference vector $\eta$, which is naturally introduced
in the method. This is an important  test which should be passed
by any proposal for an MHV diagrammatic method. 

We will suggest an off-shell continuation of the gravity MHV amplitudes 
which has precisely the effect of removing any unwanted $\eta$-dependence 
in the final result of the MHV diagram calculation,
which correctly reproduces the known expression for the four-point MHV amplitude at one loop.
Our ``experimental" prescription for the off-shell continuation, discussed in section 2,  
is based on the introduction of certain shifts  for the anti-holomorphic spinors associated to the internal (loop) legs. 
This prescription is unique and has the advantage of preserving momentum conservation
at each MHV vertex (in a sense to be fully specified  in section 2). 
The mechanism at the heart of the cancellation of $\eta$-dependence
is that of the ``box reconstruction"  found in \cite{bst},
where a generic two-mass easy box function is derived from summing
over dispersion integrals of the four cuts of the function (the
$s$- and $t$-channel cuts, and the cuts corresponding
to the two massive corners).  Each of the four terms separately contains
$\eta$-dependent terms, but these cancel out when these terms are added.  
In section 3, we apply our off-shell continuation to
calculate  in detail the four-point MHV amplitude of gravitons at one loop.
Section 4  illustrates the calculation for the case of five gravitons.  
Finally, we present our conclusions  in section 5, where  we outline the procedure
to perform a calculation with an arbitrary number
of external gravitons. Some technical details of the calculations are 
discussed in the appendices.

\section{Off-shell continuation of gravity MHV amplitudes and shifts}

The main goal of this section is to discuss (and determine) a certain
off-shell continuation  of the  MHV amplitude of gravitons
which we will use as an  MHV vertex.
We will shortly see that, compared to the Yang-Mills case,
peculiar features arise in gravity, where the expression
of the MHV amplitudes  of gravitons contains both holomorphic
and anti-holomorphic spinors.

\begin{figure}[ht]
\begin{center}
\scalebox{0.7}{\includegraphics{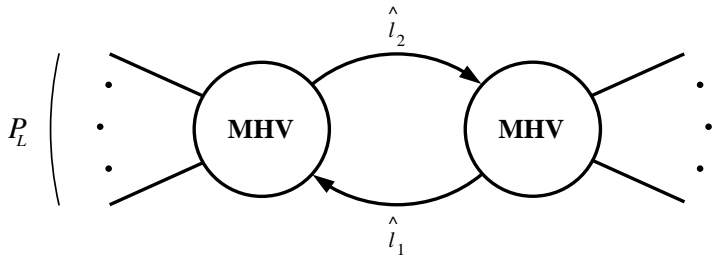}}
\end{center}
\caption{\it A generic MHV diagram contributing to the one-loop graviton MHV amplitude.
The hatted loop momenta are defined below in \eqref{ultim}.
}
\label{figure1}
\end{figure}

We start by considering the decomposition of a generic internal (possibly loop)
momentum $L$ \cite{david,bst}
which is commonly used in applications of the MHV diagram method,
\beq
\label{off}
L \ = \ l + z \eta
\ .
\eeq
Here $\eta$ is a fixed, arbitrary null vector and $z$ is
a real number; furthermore, $l^2 =  0$.
We focus on a generic MHV diagram contributing to the
one-loop MHV amplitude of gravitons, see Figure \ref{figure1}.
Using the parametrisation \eqref{off}, momentum conservation in the loop,
$L_2 - L_1 + P_L= 0$, can be rewritten as
\beq \label{prima}
P_L +l_2-l_1-z\eta \ = \ 0 \ ,
\eeq
where
\beq
\label{zdef}
z \  := \  z_1 - z_2
\ ,
\eeq
 and $P_L$ is   the sum of the momenta on the left hand side  of the diagram.

The usual CSW off-shell prescription for calculating tree-level \cite{csw} and
one-loop \cite{bst} amplitudes from MHV diagrams in Yang-Mills consists
in  decomposing  any internal  (off-shell) momentum $L$ as in \eqref{off},
and using  the holomorphic  spinor $l_\a$  associated to the null momentum
$l_{\a \da} := l_{\a}\widetilde{l}_{\da} $
in the expression of  the  MHV vertices.
In Yang-Mills, this prescription has been shown
to work for a variety of  cases at tree- \cite{v1,z1,z2,v2}
and one-loop level  \cite{bst,quig,bbst1,bbst2,ftt,pureYM}.
Moreover, Mansfield showed in \cite{Mansfield} that it arises naturally
in the framework of the  lightcone quantisation of Yang-Mills theory,
from which MHV rules are obtained via a particular
change of variables in the functional integral.%
\footnote{Very recent discussions of the specific issues arising when applying
the MHV method to the loop level  in non-supersymmetric theories
can be found in \cite{pureYM,Ettle:2007qc,bstz}.}

Using $l_1$ and $l_2$ in the expressions of the vertices in place
of the loop momenta $L_1$ and $L_2$ has the consequence of
effectively ``breaking" momentum conservation at each vertex%
\footnote{This  effective violation of momentum conservation was already observed  
and discussed in section 2 of \cite{bbst3}.}  --
the momenta which are inserted in the expression of each MHV vertex do not sum
to zero, as $l_2 - l_1 +P_L =  z \eta\neq 0$.
Interestingly, for tree-level Yang-Mills  it was shown in
\cite{ris} that momentum conservation can formally be reinstated by appropriately shifting
the anti-holomorphic spinors of the momenta of the external negative-helicity particles.
These shifts  do not affect the Parke-Taylor expressions of the MHV vertices,
as these only contain holomorphic spinors -- they are invisible.

The situation in gravity is quite different. The infinite sequence of MHV amplitudes of gravitons
was found by Berends, Giele and Kuijf  in \cite{bgk} and
is given by an expression which contains  both holomorphic and anti-holomorphic spinors
(for a number of external gravitons larger than three).
The new  formula for the $n$-point graviton scattering amplitude
found in \cite{bbst3} also contains holomorphic as well as anti-holomorphic spinors.
Thus,  it appears necessary to introduce a prescription for an off-shell continuation of
anti-holomorphic spinors $\widetilde{l}_{\da}$ related to the loop momenta.
We look for this prescription in a way which allows us to solve a potential ambiguity
which we now discuss.

We begin by observing that,  a priori, several expressions for the same
tree-level gravity MHV amplitude can be presented.
For example, different realisations of the KLT relations \cite{klt} may be used, or
different forms of the BCF recursion relations (two of which where considered in
\cite{bbst3} and \cite{cs}).
Upon making use of spinor identities and, crucially, of  momentum conservation,
one would discover that these different-looking expressions for the amplitudes
are actually identical.
However, without momentum conservation in place, these expressions are no longer equal.
We conclude that if we do not maintain momentum conservation at each MHV vertex,
we would face an ambiguity in selecting a specific form for
the graviton MHV vertex -- the expressions  obtained
by simply using the spinors $l_{i\a}$ and $\widetilde{l}_{i\da}$ obtained from
the null vectors $l_i = L_i - z_i \eta$, $i=1,2$
as in the Yang-Mills case, would in fact be different.
Not surprisingly, the  difference between any such two expressions
amounts to $\eta$-dependent terms; stated differently,
the expressions for the amplitudes na\"{i}vely continued off-shell
would present us with spurious $\eta$-dependence.
This ambiguity does not arise in the Yang-Mills case,
where there is a preferred, holomorphic expression for
the MHV amplitude of gluons, given by the Parke-Taylor formula.

We propose to resolve the  ambiguity arising in the gravity case
by resorting to certain shifts in the {\it loop} momenta,
to be determined shortly, which have the effect of
reinstating momentum conservation, in a way possibly reminiscent
of the tree-level gravity MHV rules of \cite{bdipr}.
As we shall see, these shifts determine a specific prescription
for the off-shell continuation of the spinors associated to the loop legs.

Specifically, our procedure consists in interpreting the
term $-z \eta$ in \eqref{prima} as generated by a shift  on the
{\it anti-holomorphic spinors}  of the loop momenta  in the
off-shell continuation of the MHV amplitudes.
Absorbing this extra term into the definition
of  shifted  momenta $\hat{l}_1$ and $\hat{l}_2$
allows us to preserve momentum conservation at each vertex
also off shell. Indeed, we now write momentum conservation  as
\beq
\label{consvert1}
P_L+\hat{l}_2-\hat{l}_1 \ = \ 0 \ .
\eeq
The hatted loop momenta are defined by a shift in the anti-holomorphic spinors,
\beq
\label{ultim}
\hat{l}_{1\a \da}  \ = \ l_{1\a} \hat{\widetilde{l}}_{1 \da} \ , \quad
\hat{l}_{2\a \da}  \ = \ l_{2\a} \hat{\widetilde{l}}_{2 \da}
\ .
\eeq
We find that the form of the  shifts is natural and unique.
Solving for the  anti-holomorphic spinors
$\hat{\widetilde{l}}_1$ and $\hat{\widetilde{l}}_2$, one gets%
\footnote{Notice that the off-shell prescription for the {\it holomorphic} spinors
$l_{1\a}$ and $l_{2\a}$ is the usual CSW prescription.}
\beqa
\label{shift}
\hat{\widetilde{l}}_1 \ = \ \widetilde{l}_1-z \,
\frac{\lan l_2 \eta \ran}{\lan l_1 l_2 \ran} \, \widetilde{\eta} \ , \nonumber \\
\hat{\widetilde{l}}_2 \ = \ \widetilde{l}_2-z \, \frac{\lan l_1 \eta \ran}{\lan l_1 l_2 \ran} \, \widetilde{\eta}
\ .
\eeqa
It is easy to check that the contribution of the shifts is
\beq
l_{2\a} \delta \widetilde{l}_{2\da} - l_{1\a} \delta \widetilde{l}_{1\da} \ = \
 - z \eta_{\a} \widetilde{\eta}_{\da} \ ,
\eeq
where we have used the Schouten identity
$(\lan l_1 \eta \ran \, l_{2\a} - \lan l_{2} \eta \ran \, l_{1\a} )/ \lan l_1 l_2 \ran
  =  \eta_{\a}$.

Our prescription \eqref{shift} will then consist in replacing
all the anti-holomorphic spinor variables associated to loop momenta
with corresponding shifted spinors.
For example,  the spinor bracket $[ l_2 l_1 ]$ becomes
\beq
[\hat{l}_2 \hat{l}_1] \ = \ [l_2 l_1] - 2 z \, \frac{P_L \cdot \eta}{\lan l_1 l_2 \ran}
\ .
\eeq
Notice also that
\beq
\label{slhat}
s_{\hat{l}_1 -\hat{l}_2} \ := \ (\hat{l}_2 -\hat{l}_1)^2 =
 - \lan l_1 l_2 \ran [\hat{l}_2 \hat{l}_1] \ = \ P_{L}^2
\ .
\eeq
A few comments are now in order.

\noindent
{\bf 1. }
In \cite{ris}, a derivation of tree-level MHV rules in Yang-Mills
was discussed which makes use of shifts in the momenta of {\it external} legs.
This approach was used in \cite{bdipr} where a long sought-after derivation
of tree-level gravity MHV rules was presented.
We differ from the approach of  \cite{ris} and \cite{bdipr}
in that we shift the momenta of the (off-shell) loop legs 
rather than the external momenta.
It would clearly be interesting to find a  first principle derivation of the
shifts \eqref{shift}, perhaps from an action-based approach, along the lines of
\cite{Mansfield}, as well as to relate our shifts to those employed at tree level in 
\cite{bdipr}.

\noindent
{\bf 2.}
Our procedure of shifting the loop momenta in order
to preserve momentum conservation  off shell
can also be  applied  to MHV diagrams in Yang-Mills.
Indeed, using the Parke-Taylor expression for the
MHV vertices  would result in these shifts being invisible.
We would like to point out that, in principle,
one could use different expressions
even for an MHV gluon scattering amplitude,
possibly containing anti-holomorphic spinors.
Had one chosen this second (unnecessarily complicated)
path,   our prescription \eqref{shift} for shifts in anti-holomorphic
spinors would guarantee that the non-holomorphic form of the {\it vertex}
would always boil down to the Parke-Taylor form.
Clearly, having to deal with holomorphic vertices, as in Yang-Mills, is
a great simplification. The importance of holomorphicity of the MHV amplitudes
is further  appreciated in  Mansfield's derivation  \cite{Mansfield}
of tree-level MHV rules in Yang-Mills.

In the next section we will test the ideas discussed earlier in
a one-loop calculation in $\cN=8$ supergravity, specifically
that of  a four-point MHV scattering amplitude of gravitons.
We will then consider applications to amplitudes
with arbitrary number of external particles.

\section{Four-point MHV amplitude at one loop with MHV diagrams}
\label{fourpointsection}

In this section we will rederive  the known expression for the four-point MHV
scattering amplitude of gravitons $\cM(1^- 2^- 3^+ 4^+)$ using MHV rules.  
As in the Yang-Mills case,  we will have to sum over all possible MHV diagrams, i.e.~diagrams
such that all the vertices have the MHV helicity configuration. Moreover, we will also sum
over all possible internal helicity assignments, and over the particle species which
can run in the loop.
Specifically, we will focus on $\cN=8$ supergravity, where
all the one-loop amplitudes are believed to be expressible as
sums of  box functions only  \cite{zero,zerouno,Bjerrum-Bohr:2005xx}.
In this case, the  result of \cite{Green:1982sw,dn} is
\beq
\label{target}
\cM_{\rm 1-loop}^{\cN=8} \ = \ \cM^{\rm tree} \left[ u \, F(1234)+t \, F(1243)+s \, F(1324) \right]
\ ,
\eeq
where $ \cM^{\rm tree} $ is the four-point MHV amplitude, and $F(ijkl)$
are zero-mass box functions with external, cyclically ordered null momenta
$i$, $j$, $k$ and $l$. The kinematical invariants $s$, $t$, $u$ are defined as
$s := (k_1 + k_2)^2$, $t := (k_2 + k_3)^2$, $u := (k_1 + k_3)^2 = - s - t$.
We will see in our MHV diagrams approach that  each box function appearing in
\eqref{target} will emerge by summing over appropriate dispersion integrals of
two-particle phase space integrals,  similarly to the Yang-Mills case \cite{bst}.
The result we will find is in complete agreement with the known expression
found in  \cite{Green:1982sw,dn}.

\subsection{MHV diagrams in the $s$-, $t$-, and $u$-channels}
\begin{figure}[ht]
\begin{center}
\scalebox{0.7}{\includegraphics{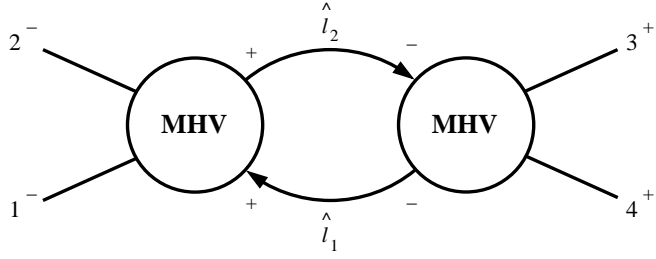}}
\end{center}
\caption{\it The $s$-channel MHV diagram.
}
\label{dueedue}
\end{figure}

We start by computing the MHV diagram in Figure \ref{dueedue}. This diagram has a nontrivial
$s$-channel cut, hence we will refer to it as to the ``$s$-channel MHV diagram".
Its expression is given by
\beqa
\label{ness}
\cM_s \ = \ \int\!  d\mu_{k_1 + k_2} \
\cM(1^- 2^- \hat{l}_2^+ -\hat{l}_1^+) \ \cM(\hat{l}_1^- -\hat{l}_2^- 3^+ 4^+)
\  .
\eeqa
The integration measure $d \mu_{P_L}$  is \cite{bst}
\beq \label{measure}
d\m_{P_L} \ = \  \frac{d^4L_1}{L_1^2+ i \varepsilon} \frac{d^4L_2}{L_2^2+ i \varepsilon} \ \delta^{(4)}(L_2-L_1+P_L)
\ ,
\eeq
where, for the specific case of  \eqref{ness}, we  have $P_L = k_1 + k_2$.
Notice the hats in \eqref{ness}, which stand for the shifts defined  in \eqref{shift}.
These shifts are such to preserve momentum conservation off shell, hence we can
use any of the (now equivalent) forms of MHV  amplitudes of gravitons
as off-shell vertices.
We choose the expression for the four-graviton MHV amplitude obtained by applying the KLT relation
\eqref{4}, thus getting
\beqa
\label{scc}
\cM(1^- 2^- \hat{l}_2^+ -\hat{l}_1^+) & =&
 - i s_{12} \, \cA(1^- 2^- l_2^+ -l_1^+)  \,  \cA(1^- 2^- -l_1^+ l_2^+)
 \ ,
 \\ \cr
 \cM(\hat{l}_1^- -\hat{l}_2^- 3^+ 4^+) &=& -i s_{\hat{l}_1 -\hat{l}_2} \, \cA(l_1^- -l_2^- 3^+ 4^+)\,
 \cA(l_1^- -l_2^- 4^+ 3^+)
  \ ,
 \eeqa
where $\cA$'s  are Yang-Mills amplitudes. We need not shift the $l$'s
appearing  inside the gauge theory amplitudes, as these are holomorphic
in the spinor variables.

Using the Parke-Taylor formula for the MHV amplitudes and the result \eqref{slhat},
the $s$-channel MHV diagram gives
\beq \label{inte}
\cM_s \ = \ - \frac{\lan 12 \ran^8}{\lan 12 \ran^2 \lan 34 \ran^2} \, s^2 \, \int\! d\m_{k_1 + k_2} \
 \, \frac{\lan l_1 l_2 \ran^4}{\lan 1 l_1 \ran \lan 2 l_1 \ran \lan 3 l_1 \ran \lan 4 l_1 \ran
\lan 1 l_2 \ran\lan 2 l_2 \ran\lan 3 l_2 \ran\lan 4 l_2 \ran} \ .
\eeq
\begin{figure}[ht]
\begin{center}
\scalebox{0.7}{\includegraphics{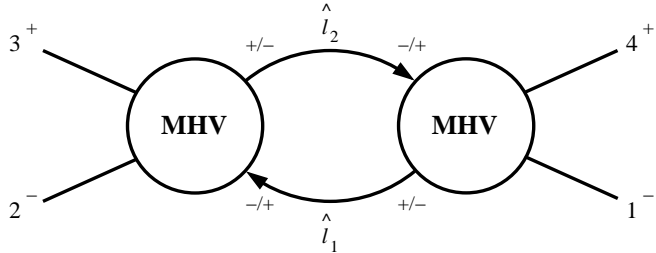}}
\end{center}
\caption{\it The $t$-channel MHV diagram. The $u$-channel diagram is obtained by 
exchanging gravitons $1^-$ and $2^-$.
}
\label{figure3}
\end{figure}

Two more MHV diagrams with a non-null two-particle cut contribute to the one-loop four-graviton amplitude, 
see  Figure \ref{figure3}.
Since these have a nontrivial $t$-channel, or $u$-channel two-particle cut,
we will call them $t$-channel,  and $u$-channel  MHV diagram, respectively.
For these diagrams, all particles in the $\cN=8$ supergravity multiplet
can run in  the loop, and moreover we will have to sum over the two possible internal
helicity  assignments.
Using supersymmetric Ward identities \cite{gpvn,mp}  it is possible to write
this sum over contributions from all particles running in the loop as the contribution
arising from a scalar loop  times a purely holomorphic quantity  $\rho_{\cN=8} $  \cite{dn},
where
\beq
\rho_{\cN=8} \ := \ \frac{\lan 1 2 \ran^8 \lan l_1 l_2 \ran^8}{(\lan 1 l_2 \ran \lan 2 l_1 \ran \lan 1 l_1 \ran \lan 2 l_2 \ran)^4}
\ .
\eeq
It is then easy to check that the results in the
$t$- and $u$-channels are exactly the same as the $s$-channel,
with the appropriate relabeling of the external legs
(apart from the overall factor $\lan 1 2 \ran^8$).
For example, in the $t$-channel we find
\beq
\label{intet}
\cM_t \ = \ - \frac{\lan 12 \ran^8}{\lan 23 \ran^2 \lan 41 \ran^2} \, t^2 \, \int\! d\mu_{k_2 + k_3} \
 \, \frac{\lan l_1 l_2 \ran^4}{\lan 1 l_1 \ran \lan 2 l_1 \ran \lan 3 l_1 \ran \lan 4 l_1 \ran
\lan 1 l_2 \ran\lan 2 l_2 \ran\lan 3 l_2 \ran\lan 4 l_2 \ran} \ .
\eeq
Making use of momentum conservation, it is immediate to see that
the prefactors of \eqref{inte} and \eqref{intet} are identical, $ s^2 / (\lan 12 \ran \lan 34 \ran)^2 =
t^2 / (\lan 23 \ran \lan 41 \ran)^2 $.

We will discuss
the specific evaluation of the $s$-channel MHV diagram
\eqref{inte} and the $t$- and $u$-channel diagrams in section \ref{calc}.
Before doing so, we would like to first write the expressions of
the remaining MHV diagrams, which have a  null  two-particle cut.

\subsection{Diagrams with null  two-particle cut}
\label{subsecnull}

In the unitarity-based approach of BDDK, diagrams with a null two-particle cut are of course irrelevant,
as  they do not have a discontinuity. However in the MHV diagram method we have to
consider them \cite{bst,bbst1,bbst2}.
As also observed in  the calculation of the gauge theory amplitudes considered in those papers,
we  will see that these diagrams give rise to contributions proportional to
dispersion integrals  of (one-mass or zero-mass) boxes in a channel with null momentum.
For generic choices of $\eta$  the contribution of these diagrams is  non-vanishing,
and  important in order to achieve the cancellation of $\eta$-dependent terms.
For specific, natural choices of $\eta$ \cite{bst}, one can see that these diagrams
actually vanish by themselves; see  appendix \ref{appendixB} for a discussion of this point.


\begin{figure}[ht]
\begin{center}
\scalebox{0.7}{\includegraphics{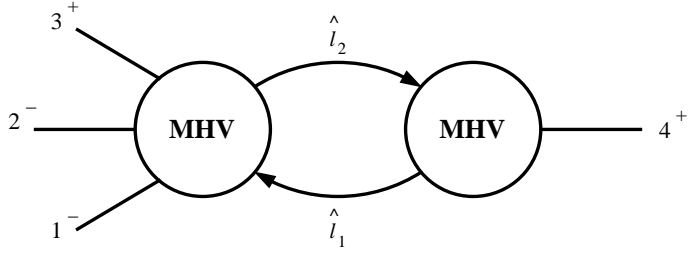}}
\end{center}
\caption{\it One of the MHV diagrams with a null two-particle cut.
}
\label{treeuno}
\end{figure}

To be specific, let us  consider the diagram with particles 1, 2 and 3 on the left,
and particle 4 on the right (see  Figure \ref{treeuno}).
The remaining three diagrams (with particle 4 replaced by particles 1, 2, and 3, respectively)
are obtained by  relabeling the external particles and summing over the
particles running in the loop, when required.

The action of the shifts \eqref{shift} allows us  to preserve momentum
conservation off shell in the form
\beq  \label{cons2}
k_1+k_2+k_3+\hat{l}_2-\hat{l}_1 \ = \ 0
\ ,
\eeq
on the left, and
\beq \label{cons3}
k_4-\hat{l}_2+\hat{l}_1 \ = \ 0 \ ,
\eeq
on the right. Equations \eqref{cons2} and \eqref{cons3} again imply that
global momentum conservation
$\sum_{i=1}^4 k_i = 0$ is also  preserved.

The expression for the diagram in Figure \ref{treeuno}  is given by
\beq
\label{ness2}
\cM_{k_4^2}  \ = \ \int\! d\mu_{k_4} \
\cM(1^- 2^- 3^+ \hat{l}_2^+ -\hat{l}_1^+) \ \cM(\hat{l}_1^- -\hat{l}_2^-  4^+)
\ .
\eeq
In order to obtain an expression for  the five-point tree-level vertex entering
\eqref{ness2},  we  apply the KLT relation \eqref{5}, whereas
for the three-point vertex we simply use \eqref{3}.  Thus, we get
\beqa
\label{pen}
\cM_{k_4^2} \  =  \  \int\! d\mu_{k_4} \ \left[
 s_{12} s_{3\hat{l}_2} \, \cA(1^- 2^- 3^+ l_2^+ -l_1^+) \cA(2^- 1^- l_2^+ 3^+ -l_1^+)
\right. \nonumber \\
\left.
\  \ \ + \,  s_{13} s_{2\hat{l}_2} \, \cA(1^- 3^+ 2^- l_2^+ -l_1^+) \cA(3^+ 1^- l_2^+ 2^- -l_1^+) \right]
\left[ \cA(l_1^- -l_2^- 4^+) \right]^2 \ ,
\eeqa
where the vector $\hat{l}_2$ is shifted.

We can now rewrite \eqref{pen} as
\beqa
\label{inte2}
\cM_{k_4^2} \ & = &  \ \frac{\lan 12 \ran^8}{\lan 12 \ran \lan 13 \ran \lan 32 \ran}
  \int\! d\mu_{k_4} \
\left[\lan 13 \ran [21] \lan 2 l_2 \ran [\hat{l}_2 3] - \lan 12 \ran
[31] \lan 3 l_2 \ran [\hat{l}_2 2] \right] \cdot  \nonumber \\
 & \cdot & \ \ \frac{\lan l_1 l_2 \ran^5}{\lan 1 l_1 \ran \lan 1 l_2 \ran \lan 2 l_1 \ran \lan 2 l_2 \ran
\lan 3 l_1 \ran\lan 3 l_2 \ran\lan 4 l_1 \ran^2\lan 4 l_2 \ran^2} \
\ .
\eeqa
Notice that apparently \eqref{inte2} contains unphysical double poles in $\lan4 l_1 \ran$ and $\lan4 l_2 \ran$,
generated  by the presence of the three-point vertex
$\left[ \cA(l_1^- -l_2^- 4^+) \right]^2$ in \eqref{pen}.
What we are going to show is that thanks to  momentum conservation --   now
always preserved in terms of the shifted momenta --  these double poles disappear.
Furthermore, we will show that the integrand
has exactly  the same form as that in \eqref{inte}, obtained from 
 diagrams with a two-particle cut.

We start by factorising out of  the integrand \eqref{inte2} the quantity
\beq
\label{Q}
Q \: = \  \frac{\lan l_1 l_2 \ran^4}{\prod_{i=1}^4 \lan i l_2 \ran \ \prod_{j=1}^4 \lan j l_1 \ran} \ .
\eeq
We are then left with
\beq
\label{senzanome}
\frac{\lan 12 \ran^8}{\lan 12 \ran \lan 13 \ran \lan 32 \ran} \,
\left[\lan 13 \ran [21] \lan 2 l_2 \ran [\hat{l}_2 3] - \lan 12 \ran
[31] \lan 3 l_2 \ran [\hat{l}_2 2] \right] \, \frac{\lan l_1 l_2 \ran}{\lan 4 l_1 \ran \lan 4 l_2 \ran}
\ .
\eeq
By using momentum conservation \eqref{cons3} on the right hand  side MHV vertex,
we can rewrite \eqref{senzanome} as
\beq
\frac{\lan 12 \ran^8}{\lan 12 \ran \lan 13 \ran \lan 32 \ran} \,
\left[\lan 13 \ran [21] [34] \frac{\lan 2 l_2 \ran}{\lan 4 l_2 \ran} -
 \lan 12 \ran
[31] [24] \frac{\lan 3 l_2 \ran}{\lan 4 l_2 \ran} \right]
\ .
\eeq
Using momentum conservation $\sum_{i=1}^4 k_i = 0$ in the form
\beq
\lan 3 l_2 \ran [31] \ = \ - \lan 4 l_2 \ran [41] - \lan 2 l_2 \ran [21]
\ ,
\eeq
we get
\beq
\label{tf}
\frac{\lan 12 \ran^8}{\lan 12 \ran \lan 13 \ran \lan 32 \ran} \,
\left[ \lan 12 \ran [24] [41] + ( \lan 13 \ran [34] + \lan 12 \ran [24]) [21]
\frac{\lan 2 l_2 \ran}{\lan 4 l_2 \ran} \right] \ = \
\frac{\lan 12 \ran^8}{\lan 12 \ran \lan 13 \ran \lan 32 \ran}  \lan 12 \ran [24] [41]
\ .
\eeq
The surprise is that the coefficient \eqref{tf} is actually
the negative of the prefactor which multiplies the integral in  the expression \eqref{inte}
for  the MHV diagrams corresponding to  the $s$-channel.
We can thus   rewrite \eqref{inte2} as
\beq
\label{K2}
\cM_{k_4^2} \ = \  \frac{\lan 12 \ran^8}{\lan 12 \ran^2 \lan 34 \ran^2} \, s^2 \, \int\! d\mu_{k_4} \
 \, \frac{\lan l_1 l_2 \ran^4}{\lan 1 l_1 \ran \lan 2 l_1 \ran \lan 3 l_1 \ran \lan 4 l_1 \ran
\lan 1 l_2 \ran\lan 2 l_2 \ran\lan 3 l_2 \ran\lan 4 l_2 \ran}
\ ,
\eeq
which is the opposite of the right hand side of  \eqref{inte} -- except for the integration measure
$d\m_{k_4^2}$
appearing in \eqref{K2}, which  is different from that in \eqref{inte}
(as the momentum flowing in the cut is different).
As we shall see in the next section, the  relative minus sign found in \eqref{K2} compared to \eqref{inte}
is precisely needed in order to reconstruct box functions from summing dispersive integrals
(see \eqref{sumimp}),  one for each cut, as it was found in \cite{bst}.


\subsection{Explicit evaluation of the one-loop MHV diagrams}
\label{calc}

In the last sections we have encountered a peculiarity of the gravity calculation,
namely the fact that  the expression for the integrand of each MHV diagram contributing
to the four-point graviton MHV amplitude turns out to be the same -- compare, for example,
\eqref{inte}, \eqref{intet}, \eqref{K2}, which correspond to the $s$-, $t$-, and $k_4^2$-channel MHV diagram,  respectively.
Therefore we will focus on  the expression of a generic contribution of these MHV diagrams,
for example from \eqref{inte},
\beq \label{intepl}
\cM \ = \ - \frac{\lan 12 \ran^8}{\lan 12 \ran^2 \lan 34 \ran^2} \, s^2 \, \int\! d\m_{P_L} \
 \, \frac{\lan l_1 l_2 \ran^4}{\lan 1 l_1 \ran \lan 2 l_1 \ran \lan 3 l_1 \ran \lan 4 l_1 \ran
\lan 1 l_2 \ran\lan 2 l_2 \ran\lan 3 l_2 \ran\lan 4 l_2 \ran} \ ,
\eeq
and perform the relevant phase space and dispersion integrals.

In order to evaluate  \eqref{intepl}, we need to perform the PV reduction
of the phase-space integral  of  the quantity $Q$ defined in \eqref{Q}.
To carry out this reduction efficiently,
we use the trick of performing certain auxiliary shifts,
which allow us to decompose  \eqref{Q}  in partial  fractions.
Each term produced in this way will then have a very simple PV reduction.

Firstly,  we write $Q$ as
\beq
Q \ := \lan l_1 l_2 \ran^4\,  \ X \,  Y\,
\ ,
\eeq
where
\beqa
\label{X}
X& = & \frac{1}{\prod_{i=1}^4 \lan i l_2 \ran} \ ,
\\
Y&=&\frac{1}{\prod_{j=1}^4 \lan j l_1 \ran}
\ ,
\label{Y}
\eeqa
and perform the following auxiliary shift
\beq
\hat{\lambda}_{l_2} \ = \ \lambda_{l_2}+ \omega \lambda_{l_1}
\ ,
\eeq
on the quantity  $X$ in \eqref{X} (we will later apply the same procedure on  $Y$).
We call $\hat{X}$ the corresponding shifted quantity,
\beq
\hat{X}\ =\ \frac{1}{\prod_{i=1}^4 (\lan i l_2 \ran + \omega \lan i l_1 \ran)} \ .
\eeq
Next,  we decompose $\hat{X}$
in partial fractions, and finally set  $\omega=0$.
After using the Schouten  identity,  we find that $X$ can be recast as
\beq
X \ = \ \frac{1}{\lan l_1 l_2 \ran^3} \sum_{i=1}^4 \, \frac{\lan i l_1 \ran^3}{\prod_{m \neq i} \lan i m \ran} \,
\frac{1}{\lan i l_2 \ran} \ .
\eeq
One can proceed in a similar way for $Y$ defined in \eqref{Y},
and, in conclusion,   \eqref{Q}  is re-expressed as
\beq
\label{Q1}
Q \ = \ \sum_{i,j=1}^4 \, \frac{1}{\prod_{m \neq i} \lan i m \ran} \,
\frac{1}{\prod_{l \neq j} \lan j l \ran} \, \frac{1}{\lan l_1 l_2 \ran^2} \,
\frac{\lan i l_1 \ran^3 \, \lan j l_2 \ran^3}{\lan i l_2 \ran \, \lan j l_1 \ran}
\ .
\eeq
We now set
\beq \label{Q2}
Q \ = \   \sum_{i,j=1}^4 \, \frac{1}{\prod_{m \neq i} \lan i m \ran} \,
\frac{1}{\prod_{l \neq j} \lan j l \ran} \  K \ ,
\eeq
where
\beq
K \ := \  \frac{1}{\lan l_1 l_2 \ran^2} \,
\frac{\lan i l_1 \ran^3 \, \lan j l_2 \ran^3}{\lan i l_2 \ran \, \lan j l_1 \ran}
\ ,
\eeq
and substitute the Schouten identity for the factor $(\lan i l_1 \ran \lan j l_2 \ran)^2$ in $K$.
By multiplying for
appropriate anti-holomorphic inner products (of  unshifted spinors), we are able to
reduce $K$  to the sum of three  terms as follows:
\beq
\label{K}
K \ = \ \frac{\lan i | \, l_2 P_{L;z} \, | i \ran \lan  j | \, l_2 P_{L;z} \, | j \ran}{(P_{L;z}^2)^2}
+ 2 \lan ij \ran \frac{\lan j | \, l_2 P_{L;z} \, | i \ran}{P_{L;z}^2} +
\lan ij \ran^2 R(ji)
\ ,
\eeq
where
\beq
P_{L;z} \ := \ P_{L} - z \eta \ ,
\eeq
and $z$ is  defined in \eqref{zdef}.
The first term in \eqref{K}
 gives two-tensor bubble integrals, the second  linear bubbles, and the  third term
generates the usual $R$-function, familiar from the Yang-Mills case. This is defined by
\beq
\label{erre}
R(ji) \ = \ \frac{\lan j l_2 \ran \lan i l_1 \ran}{\lan j l_1 \ran \lan i l_2 \ran}
\ .
\eeq
We can then decompose the  $R$ function as
\beqa
\label{erre2}
R(ji) \ & = & \ \frac{2 \left[ (l_1 j)(l_2 i)+(l_1 i)(l_2 j)-(l_1 l_2)(ij) \right]}{(l_1-j)^2 (l_2+j)^2} \
\nonumber \\
 \ \ \ & = & \   - 1  +\frac{1}{2} \left[ \frac{P_{L;z} i}{l_2 i} - \frac{P_{L;z} j}{l_1 j} \right] +
\frac{2(i P_{L;z})(j P_{L;z})-P_{L;z}^2 (ij)}{4 (l_2 i)(l_1 j)}
\ .
\eeqa
The phase-space integral of the first term
on the right hand side of \eqref{erre2} corresponds to  a scalar bubble,
whereas the second and the third one correspond to
triangles; finally, the phase-space integral of the last term in
\eqref{erre2}  gives rise  to a box function.
The last term is usually called
$R^{\rm eff}(ji)$,
\beq
\label{erreff}
R^{\rm eff}(ji) \ := \  \frac{N(P_{L;z})}{(l_1-j)^2 \, (l_2+i)^2}
\ ,
\eeq
where
\beq
N(P_{L;z}) \ := \ -2 (i  P_{L;z}) \, (j  P_{L;z}) + P_{L;z}^2 (i  j)
\ .
\eeq
We  now show the cancellation of bubbles and triangles,  which leaves us
just with box functions.

To start with,  we pick   all contributions to  (the phase-space integral of) \eqref{Q2}
corresponding to   scalar, linear and two-tensor bubbles, which we identify using \eqref{K}.
These are
given by
\beq \label{Q3}
Q_{\rm bubbles} \ = \   \sum_{i,j=1}^4 \, \frac{1}{\prod_{m \neq i} \lan i m \ran} \,
\frac{1}{\prod_{l \neq j} \lan j l \ran} \  \left[ \frac{\lan i | \, l_2 P_{L;z} \, | i \ran \lan  j | \,
 l_2 P_{L;z} \, | j \ran}{(P_{L;z}^2)^2}
+ 2 \lan ij \ran \frac{\lan j | \, l_2 P_{L;z} \, | i \ran}{P_{L;z}^2} -\lan ij \ran^2 \right] \ .
\eeq
Explicitly, the phase-space integrals of  linear and two-tensor bubbles are given by%
\footnote{Up to a common constant, which will not be needed in the following.}
\beq
I^{\mu} \ = \ \int\! d\textrm{LIPS}(l_2,-l_1;P_{L;z}) \ l_2^{\mu} \ = \ - \frac{1}{2} \, P_{L;z}^{\mu} \ ,
\eeq
and
\beq
I^{\mu \nu} \ = \ \int\! \  d\textrm{LIPS}(l_2,-l_1;P_{L,z}) \ l_2^{\mu} \, l_2^{\nu} \ = \
\frac{1}{3}\left[ P_{L;z}^{\mu} P_{L;z}^{\nu} - \frac{1}{4} \eta^{\mu \nu} P_{L;z}^2 \right]  \ .
\eeq
Thus,  we find that the bubble contributions arising from \eqref{Q3}
give a result proportional to
\beqa \label{C}
C  & = &  \sum_{i,j=1}^4 \, \frac{\lan ij \ran^2}{\prod_{m \neq i} \lan i m \ran
\prod_{l \neq j} \lan j l \ran} \,  \ .
\eeqa
Using  the Schouten identity, it is immediate to show that
$C  = 0 $.
We remark that the previous expression vanishes also for a fixed value of $i$.

We now move on to consider the triangle contributions.
From \eqref{Q2} and \eqref{erre2}, we get
\beq
\label{qtr}
Q_{\rm triangles} \ = \ \sum_{i,j=1}^4 \, \frac{1}{\prod_{m \neq i} \lan i m \ran} \,
\frac{1}{\prod_{l \neq j} \lan j l \ran} \ \frac{\lan ij \ran^2}{2} \left[ \frac{P_{L;z} i}{l_2 i} - \frac{P_{L;z} j}{l_1 j} \right] \ .
\eeq
We observe  that the combination
\beq
\int\! d{\rm LIPS} \ \left[ \frac{P_{L;z} j}{l_1 j} -
\frac{P_{L;z} i}{l_2 i} \right] \ = \ - \frac{4 \pi \lambda}{\epsilon} \ ,
\eeq
is independent of $i$ and $j$  \cite{bbst1},  hence  we can bring the corresponding term in \eqref{qtr}
outside the summation, obtaining again a contribution proportional to the
coefficient  \eqref{C}, which vanishes; this proves the cancellation of  triangles.
We conclude that each one-loop MHV diagram is written just in terms of box functions,
and is explicitly given by
\beq \label{result}
\cM \ = \ - \frac{\lan 12 \ran^8}{\lan 12 \ran^2 \lan 34 \ran^2} \, s^2 \, \int\! d\mu_{P_L} \,
 \, \sum_{i \neq j} \, \frac{ \lan i j \ran^2}{\prod_{m \neq i} \lan i m \ran \, \prod_{l \neq j} \lan j l \ran}
 \frac{N(P_{L;z})}{(l_1-j)^2 \, (l_2+i)^2}
 \ .
\eeq
We remind that $P_L$ is the sum of the (outgoing) momenta in the
left hand side  MHV vertex. To get the full amplitude at one loop we will then
have to sum over all possible MHV diagrams.

The next task consists in performing the loop integration.
To do this, we follow steps similar  to those discussed in \cite{bst},
namely:

{\bf 1.}
We rewrite the integration measure as the product of a Lorentz-invariant phase space measure
and an integration over the $z$-variables (one for each loop momentum) introduced by
the off-shell continuation,%
\footnote{In this and following formulae, the appropriate $i \varepsilon$ prescriptions
are understood. These have been  extensively discussed in section 5 of \cite{ftt}.}
\beq
 d\mu_{P_L} \ := \ \frac{d^4L_1}{L_1^2} \frac{d^4L_2}{L_2^2} \ \delta^{(4)}(L_2-L_1+P_L) \ = \
\frac{dz_1}{z_1} \frac{dz_2}{z_2} \, d{\rm LIPS}(l_2,-l_1;P_{L;z}) \ .
\eeq

{\bf 2.}
We change variables from $(z_1,z_2)$ to $(z,z^{\prime})$, where $z^{\prime}  :=  z_1+z_2$
and $z$ is defined in \eqref{zdef},
and perform a trivial contour integration over $z^{\prime}$.

\begin{figure}[ht]
\begin{center}
\scalebox{0.7}{\includegraphics{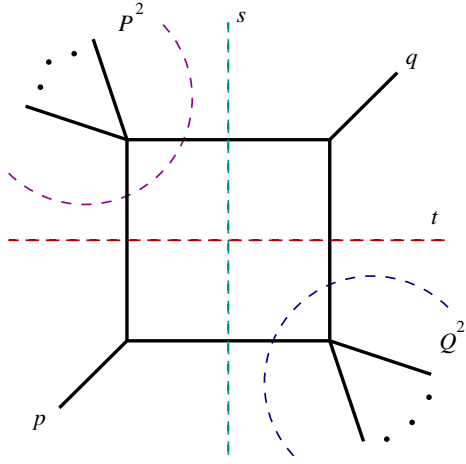}}
\end{center}
\caption{\it A generic two-mass easy box function. $p$ and $q$ are the massless legs, $P$ and $Q$
the massive ones, and
$s:=(P+p)^2$, $t:=(P+q)^2$. }
\label{boxgen}
\end{figure}

{\bf 3. }
We use dimensional regularisation on the phase-space integral of the boxes,
\beq
\cP \ = \ \int\! d^D {\rm{LIPS}}(l_2,-l_1;P_{L}) \ \frac{N(P_{L})}{(l_1-j)^2 \, (l_2+i)^2} \ .
\eeq
This evaluates to all orders in $\epsilon$ to
\beq
\label{ip}
\cP \ = \ \frac{\pi^{\frac{3}{2}-\epsilon}}{\Gamma(\frac{1}{2}-\epsilon)}
\, \frac{1}{\epsilon} \,  \left| \frac{P_{L}^2}{4} \right|^{- \epsilon} \, {}_2F_{1}(1, - \epsilon, 1- \epsilon, a P_{L}^2) \ ,
\eeq
where
\beq
\label{adef}
a \ := \ \frac{ P^2+Q^2-s-t}{P^2Q^2-st}
\ .
\eeq
The phase space integral in \eqref{ip} is computing a particular discontinuity of
the box diagram represented in in Figure \ref{boxgen}, with $p=i$ and $q = j$, where
the cut momentum is $P_L$.

{\bf 4.}
We  perform the final $z$-integral by defining the new variable
\beq
s^{\prime} \ := \ P_{L;z}^2 \ = \ P_L^2 - 2 z P_L\cdot \eta \ .
\eeq
One notices that \cite{bst}
\beq
\frac{dz}{z} \ := \ \frac{ds^{\prime}}{s^{\prime}-P_L^2} \  ,
\eeq
hence the $z$-integral  leads to a dispersion integral  in the
$P_L^2$-channel.
At this point we select a specific value for $\eta$,
namely we  choose it  to be equal to the momentum of
particles $j$ or $i$.%
\footnote{These natural choices of $\eta$, discussed in section 5 of \cite{bst},
are reviewed in appendix \ref{appendixB}.}
Specifically, performing  the phase-space integration and the dispersive integral
for a box in the $P_L^2$-channel, we get
\beqa
\label{tagl}
\int\! d\m_{P_L}  \,
\frac{N(P_{L;z})}{(l_1-j)^2 \, (l_2+i)^2} & = &  - \frac{c_{\Gamma}}{\epsilon^2}
\, (-P_L^2)^{- \epsilon} \,
{}_2F_{1}(1, - \epsilon, 1- \epsilon, a P_L^2)
\\ \nonumber \cr
& :=& F_{P_L^2} (p, P, q, Q)
\ ,
\eeqa
where
\beq
\label{cgamma}
c_\Gamma \ := \ {\Gamma (1 + \e) \Gamma^2 ( 1 -  \e) \over (4\pi)^{2- \e}
\Gamma (1 - 2 \e)}
\ .
\eeq
The subscript $P_L$ refers to the dispersive channel in which
\eqref{tagl} is evaluated; the arguments of $F_{P_L^2}$
correspond  to the ordering of the  external legs of the box function.

We can rewrite  \eqref{result}  as
\beq
\cM \ = \ - 2\, \frac{\lan 12 \ran^8}{\lan 12 \ran^2 \lan 34 \ran^2} \, s^2 \, \int\! d\mu_{P_L}  \,
 \, \sum_{i < j} \, \frac{ \lan i j \ran^2}{\prod_{m \neq i} \lan i m \ran \, \prod_{l \neq j} \lan j l \ran}
 \frac{N(P_{L;z})}{(l_1-j)^2 \, (l_2+i)^2} \ ,
\eeq
or, in terms of the $R^{\rm eff}$ functions introduced in
\eqref{erreff},
\beqa
\nonumber
\cM  &=  & - 2\, \frac{\lan 12 \ran^8}{\lan 12 \ran^2 \lan 34 \ran^2} \, s^2 \, \int\! d\mu_{P_L}  \,
 \, \bigg[ \frac{R^{\rm eff}(13)+R^{\rm eff}(24)}{\lan 1 2 \ran \lan 1 4 \ran \lan 3 2  \ran \lan 3 4 \ran}
+\frac{R^{\rm eff}(23)+R^{\rm eff}(14)}{\lan 1 2 \ran \lan 1 3 \ran \lan 4 2  \ran \lan 4 3 \ran}
\\  \cr
&  &\hspace{3.6cm}  + \ \frac{R^{\rm eff}(12)+R^{\rm eff}(34)}{\lan 13 \ran \lan 14 \ran \lan 23 \ran \lan 24 \ran} \bigg]
 \ .
\label{risultato}
\eeqa
For the sake of definiteness, we now specify the PV reduction we have performed
to the  $s$-channel MHV diagram  ($P_L \ = \ k_1+k_2$),  and
analyse  in detail the contributions to the different box functions.
In this case, the first two $R$-functions contribute
to  the box $F(1234)$,  and the second two to the box $F(1243)$.
Specifically, from these terms we obtain
\beq
\label{pr1}
\cM^{\rm tree} \left[ u \, F_s (1234) \, + \, t \, F_s (1243) \right]
\ ,
\eeq
where the subscript indicates the channel in which the dispersion integral is performed
($s := s_{12}$),  and
\beq
\cM^{\rm tree} \ := \ \frac{\lan 1 2 \ran^7 \, [12]}{\lan 1 3 \ran \lan 1 4 \ran
\lan 2 3 \ran \lan 2 4 \ran \lan 3 4 \ran^2} \
\eeq
is the tree-level four-graviton MHV scattering amplitude.

The last two terms in \eqref{risultato} give
a contribution to particular box diagrams where one of the
external legs happens to have a vanishing momentum.
\begin{figure}[ht]
\begin{center}
\scalebox{0.7}{\includegraphics{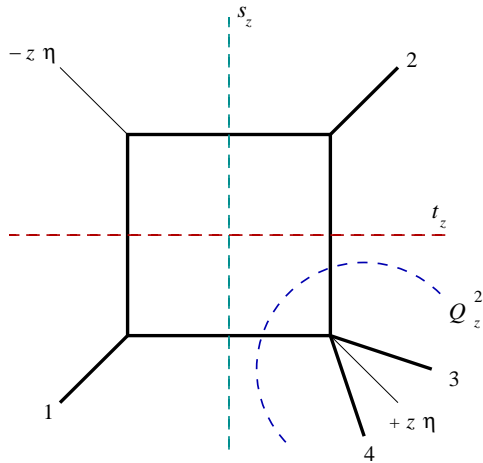}}
\end{center}
\caption{\it  Cut-box function, where -- before dispersive integration --
one of the external legs has a momentum proportional to $z \eta$. }
\label{strano}
\end{figure}
In principle,  these  boxes  are reconstructed, as all the others,
by summing over dispersion integrals in
their cuts (note that in this case there is one cut missing,
corresponding to the $\eta^2$-{\rm channel}).
However, one can see that these box diagrams give a vanishing contribution
already at the level of phase space integrals,
when $\eta$ is chosen,  for each box, in exactly the same way as in the
Yang-Mills calculation of   \cite{bst}.
For example, consider  the box diagram in Figure \ref{strano}, for which these
natural choices are $\eta = k_1$ or $\eta = k_2$.
Prior to the dispersive integration, this box has three nontrivial cuts: $s_z = (k_1 - z \eta)^2$,
$t_z = (k_2 - z \eta)^2$, and $Q_z^2 = (k_3 + k_4 + z \eta)^2$.
Using \eqref{ip} to perform the phase space integrals,
one encounters two distinct cases:
either the quantity  $a P_{L;z}^2$ is finite but $P_{L;z}^2\to 0$ ($P_{L;z} $ is the momentum
flowing in the cut); or $a P_{L;z}^2 \to \infty $.
It is then easy to see that  in both cases the corresponding contribution vanishes.%
 \footnote{In the second case, we make use of the identity
 ${}_2F_1(1,-\epsilon,1-\epsilon,z) \ = \
 (1-z)^{\epsilon} {}_2F_1\left(-\epsilon,-\epsilon,1-\epsilon,\frac{-z}{1-z}\right)$.
}
The conclusion is that such boxes   can be discarded altogether.
For the same reason these diagrams were discarded in the Yang-Mills case.

Next, we consider the $t$-channel  MHV diagram. In this case the second term in
\eqref{risultato} gives contribution to vanishing boxes like that  depicted
in Figure \ref{strano}, the first and last terms instead give the contribution:
\beqa
\label{pr2}
\cM^{\rm tree} \Big[ u \, F_t(1234) \, + \, s \, F_t(1324) \Big]
 \ .
\eeqa
Similarly, for the $u$-channel we obtain:
\beqa
\cM^{\rm tree} \Big[ s \, F_u(1324) \, + \, t \, F_u(1243) \Big] \ .
\label{pr3}
\eeqa
Again the subscript indicates the channel in which the dispersion integral is performed
($t:=s_{23}$ and $u:=s_{13}$).

As in the Yang-Mills case, we have to sum over all possible MHV diagrams.
In particular, we will also have to include the $k_1^2$-, $k_2^2$-, $k_3^2$- and $k_4^2$-channel
MHV diagrams.
In section \ref{subsecnull} we have seen that, prior to the phase space and dispersive integration,
these diagrams produce expressions  identical up to a sign to those in the $s$-, $t$-, and $u$-channels.
Hence they will give rise to dispersion integrals of the same cut-boxes found
in those channels, this time  in their $P^2$- and $Q^2$-cuts.
They appear  with the same coefficient, but  opposite sign.
We can thus collect dispersive integrals in different channels of the same box function,
which appear with the same coefficient, and
use the result proven in \cite{bst}
\beq
\label{sumimp}
F \ = \ F_s + F_t -F_{P^2}-F_{Q^2} \ ,
\eeq
in order to reconstruct each box function from
the four dispersion integrals in its $s$-, $t$-, $P^2$- and $Q^2$- channels.%
\footnote{Notice that in \eqref{sumimp}, the subscript refers to the channels
of the box function itself (which are different for each box).
For instance,  the $s$-channel ($t$-channel)
of the box $F(1324)$ is $s_{13}$ ($s_{23}$). }
For completeness, we quote from \cite{ftt} the all orders in $\epsilon$
expression for  a generic two-mass easy 
box function,  
\beqa
\nonumber
&& \hspace{-0.7cm}
F  =
-{ c_{\G} \over \e^2}
\left[
 \Big( {-s \over \mu^2} \Big)^{-\e} \,
\mbox{}_{2}F_1 \left( 1, -\e, 1- \e, as \right)
\, + \,
\Big( {-t \over \mu^2} \Big)^{-\e}
\mbox{}_{2}F_1 \left( 1, -\e, 1- \e, at \right)
\right.
\\  [6pt]\cr
&& \! - \,
\left.\Big( {-P^2 \over \mu^2} \Big)^{-\e}
\,
\mbox{}_{2}F_1 \left( 1, -\e, 1- \e, aP^2 \right)
\, - \,
 \Big( {-Q^2 \over \mu^2} \Big)^{-\e} \,
\mbox{}_{2}F_1 \left( 1, -\e, 1- \e, a Q^2 \right)
\right]
\, , 
\eeqa
where $c_\Gamma$ is defined in  \eqref{cgamma}.

\begin{figure}[ht]
\begin{center}
\scalebox{0.7}{\includegraphics{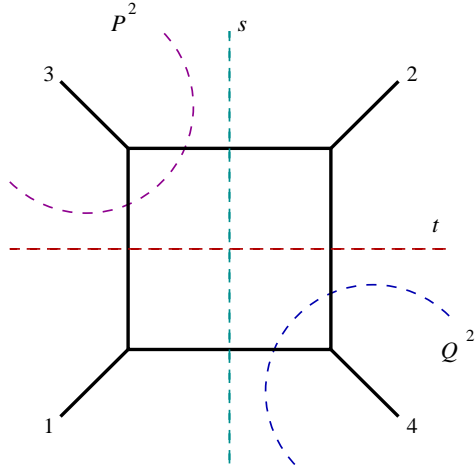}}
\end{center}
\caption{\it The box function $F(1324)$, appearing in the four-point amplitude
\eqref{rfn}. }
\label{1324}
\end{figure}

As an example, we discuss in more detail how the box
$F(1324)$ (depicted in Figure \ref{1324})  is reconstructed.
Due to the degeneracy related to the particular case of four particles,
 both the $R$-functions $R(12)$ and $R(34)$ give contribution to this box
 (see the third term  in the result \eqref{risultato}).%
 \footnote{This box is reconstructed as a  two-mass easy box with massless legs
given by the entries of the $R$-function; in the specific four-particle case, the
massive legs of the two-mass easy function are, of course, also massless.}
Let us focus on the contribution from the function $R(12)$, corresponding to the box
in Figure \ref{1324}.
This box function gets contributions from MHV diagrams in the channels
$u = s_{13}$, $t = s_{32}$, $k_3^2$ and $k_4^2$. They all appear with the same coefficient,
given by the third term in \eqref{risultato}, the last two contributions having opposite sign,
as shown (we note that for all the others diagrams this term in the result gives
contribution to vanishing boxes, as the one  in Figure \ref{strano}).
These four contributions to the box $F(1324)$ correspond to 
its cuts in the  $s  =  s_{13}${\mbox{-,}}  $t = s_{32}${\mbox{-,}}  $P^2 = k_3^2$-
and $Q^2 \!= \!k_4^2$-channels. By summing over these four dispersion integrals
using \eqref{sumimp}, we immediately reconstruct the box function $F(1324)$,
which appear with a coefficient
\beq
\cM^{\rm tree}(1^-2^-3^+4^+) \,   s \, F(1324) \ .
\eeq
This procedure can be applied in an identical fashion to reconstruct the other box functions.
Summing over the contributions from all the different channels,
and using \eqref{sumimp} to reconstruct all the box functions
we arrive at the final result
\beq
\label{rfn}
\cM^{1-{\rm loop}}(1^-2^-3^+4^+) \ = \ \cM^{\rm tree}(1^-2^-3^+4^+) \, \left[ \, u \, F(1234) \,
+ \, t \, F(1243) \, + \, s \, F(1324) \, \right]
\ .
\eeq
This is in  complete agreement with the result of \cite{dn}  found using the unitarity-based method.

\section{Five-point amplitudes}
\label{5ptampl}

We would like to discuss how the previous calculations can be
extended to the case of scattering amplitudes with more
than four particles.
To be specific, we  consider the  five-point
MHV amplitude of gravitons  $\cM(1^- 2^- 3^+ 4^+5^+)$.
Clearly, increasing the number of external particles  leads
to an increase in  the algebraic  complexity of the problem.
However, the same basic procedure discussed in
the four-particle case can be applied; in particular,
we observe that the shifts \eqref{shift}
can be used  for any number of external particles.
This set of shifts allows one
to use any on-shell technique of reduction of
the integrand. In appendix \ref{appa} we propose a reduction technique
alternative to that used in this and in section \ref{fourpointsection},
which can  easily be applied to the case of an arbitrary
number of external particles.

\begin{figure}[ht]
\begin{center}
\scalebox{0.7}{\includegraphics{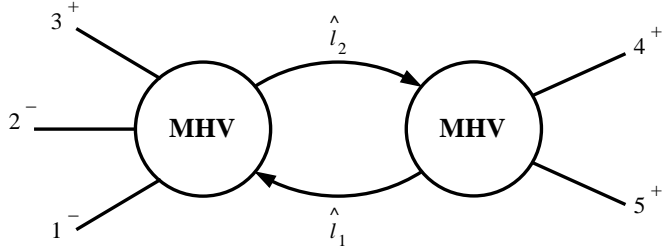}}
\end{center}
\caption{\it MHV diagram contributing to the five-point MHV amplitude discussed in the text.}
\label{treedue}
\end{figure}

We now consider the MHV diagrams contributing to the five-particle
MHV amplitude.
We start by computing the MHV diagrams
which have a non-null two-particle cut.
Firstly, consider the diagram pictured in Figure \ref{treedue}.
Its expression  is given by
\beqa
\label{fiveone}
\cM^{1-{\rm loop}}_{(123)(45)} \ = \ \int\!  d\mu_{P_{123}} \
\cM(1^- 2^- 3^+l_2^+ -l_1^+) \ \cM(l_1^- -l_2^-  4^+5^+)
\  ,
\eeqa
where $d\mu_{P_L}$ is given by \eqref{measure} and $P_{123} := k_1 + k_2 + k_3$.
We make use of  the off-shell continuation for the
anti-holomorphic spinors of the loop momenta given by  \eqref{shift},
which guarantees momentum conservation off shell --
irrespectively of the number of the particles in the vertex,
as the shifts act only on the two loop legs.

In order to evaluate \eqref{fiveone}, we need expressions
for the four- and five-point tree-level gravity MHV vertices;
these can be obtained by  using the KLT relations \eqref{4} and \eqref{5}.
Thus, we find
\beqa
\label{plan}
\cM(1^- 2^- 3^+\hat{l}_2^+ -\hat{l}_1^+) &= & i \, s_{12} s_{3 \hat{l}_2} \cA(1^- 2^- 3^+l_2^+ -l_1^+)
\,  \cA( 2^- 1^- l_2^+ 3^+ -l_1^+)
\nonumber \\
& + &  i \, s_{13} s_{2 \hat{l}_2} \cA(1^-  3^+2^-l_2^+ -l_1^+)
\cA( 3^+1^- l_2^+2^- -l_1^+) \ ,
\\ \cr
 \cM(\hat{l}_1^- -\hat{l}_2^-  4^+5^+) & = & -i \, s_{\hat{l}_1 -\hat{l}_2} \cA(l_1^- -l_2^-  4^+5^+)
\cA(l_1^- -l_2^-  5^+4^+) \ ,
\eeqa
where  $\cA$ are Yang-Mills amplitudes.
Plugging the Parke-Taylor formula for the
Yang-Mills MHV amplitudes appearing in \eqref{plan},
we get
\beqa
\label{integr}
\cM^{1-{\rm loop}}_{(123)(45)}\ & = &  \
\frac{\lan 12 \ran^8}{\lan 12 \ran \lan 13 \ran \lan 23 \ran \lan 45 \ran^2}
\int\! d\mu_{P_{123}}  \
\ s_{\hat{l}_1 -\hat{l}_2} \,
\left[\lan 13 \ran [21] \lan 2 l_2 \ran [\hat{l}_2 3] - \lan 12 \ran
[31] \lan 3 l_2 \ran [\hat{l}_2 2] \right] \cdot  \nonumber \\
 & \cdot & \ \
 \frac{\lan l_1 l_2 \ran^5}{\lan 1 l_1 \ran \lan 1 l_2 \ran \lan 2 l_1 \ran \lan 2 l_2 \ran
\lan 3 l_1 \ran\lan 3 l_2 \ran\lan 4 l_1 \ran \lan 4 l_2 \ran \lan 5 l_1 \ran \lan 5 l_2 \ran} \
\ .
\eeqa
With shifted spinors defined as in \eqref{shift},
momentum conservation is expressed as
\beq
k_1+ k_2+k_3+\hat{l}_2-\hat{l}_1 \ = \ 0 \ .
\eeq
This allows us to rewrite
\beq
\label{gc}
\lan l_1 l_2 \ran [\hat{l}_2 3] \ = \ - \lan l_1 1 \ran [1 3]- \lan l_1 2 \ran [2 3] \ ,
\eeq
and similarly  for the term in the first line of \eqref{integr} containing $[\hat{l}_2 2]$.
As in \eqref{slhat}, we can also write
$s_{\hat{l}_1 -\hat{l}_2}  = P_{L}^2 = P_{123}^2$.
Next, using relations such as \eqref{gc},
the dependence on the shifted  momenta can be completely eliminated.
Each of the four terms generated in this way  will be of the same form
as \eqref{inte}, but now with different labels of the particles.
\eqref{integr} then becomes,
\beqa
\cM^{1-{\rm loop}}_{(123)(45)} &  =  & \frac{\lan 12 \ran^8}{\lan 23 \ran \lan 45 \ran^2}
   \int\!   d\m_{P_{123}} \
   P_{123}^2   \left[ \frac{[21]}{\lan 12 \ran}   \left( [13]
\, Q_{i=1,3,4,5;j=2,3,4,5}
+ [23] \, Q_{i,j=1,3,4,5}
\right)  \right.  \nonumber \\
&&   \qquad \qquad \qquad \qquad \quad  \ \
 \left.   + \  \frac{[31]}{\lan 13 \ran} \big( [21] \, Q_{i=1,2,4,5;j=2,3,4,5}
+[23] \, Q_{i,j=1,2,4,5} \big)  \right]\ ,
\nonumber \\
\label{gat}
\eeqa
where, similarly to \eqref{Q}, the $Q$ functions are defined as
\beq
Q \ = \  \frac{\lan l_1 l_2 \ran^4}{\prod_{i} \lan i l_2 \ran \ \prod_{j} \lan j l_1 \ran} \ .
\eeq
Next,  we  decompose the integrand in \eqref{gat}  in  partial fractions,
in order to allow for a simple PV reduction,
as done earlier in the four-particle case.
It is easy to see that the outcome of this procedure
is a  sum of four terms, each of which  has the same form
as \eqref{Q1}. Specifically, the box functions contributions is
\beqa
\left.
\cM_{(123)(45)}\right|_{\rm box}&=& \frac{\lan 12 \ran^8}{\lan 23 \ran \lan 45 \ran^2}
P_{123}^2 \,
   \int\!   d\m_{P_{123}}     \left[ \frac{[21][13]}{\lan 12 \ran}  A_{i=1,3,4,5;j=2,3,4,5}+
\frac{[21][23]}{\lan 12 \ran}
  A_{i,j=1,3,4,5}
\right. \nonumber \\
&&   \qquad \qquad \qquad \qquad \quad  \ \,
\left.  +  \, \frac{[31][21]}{\lan 13 \ran}  A_{i=1,2,4,5;j=2,3,4,5}+
\frac{[31][23]}{\lan 13 \ran} \, A_{i,j=1,2,4,5} \right]
 \nonumber \\
 \label{5p}
\eeqa
where we have defined%
\footnote{This function is nothing but the integrand of  \eqref{result}.}
\beq
\label{afucts}
 A \ := \ \sum_{i,j} \, \frac{ \lan i j \ran^2}{\prod_{m \neq i} \lan i m \ran \,
 \prod_{l \neq j} \lan j l \ran}
 \frac{N(P_{L;z})}{(l_1-j)^2 \, (l_2+i)^2}  \ .
\eeq

\begin{figure}[ht]
\begin{center}
\scalebox{0.7}{\includegraphics{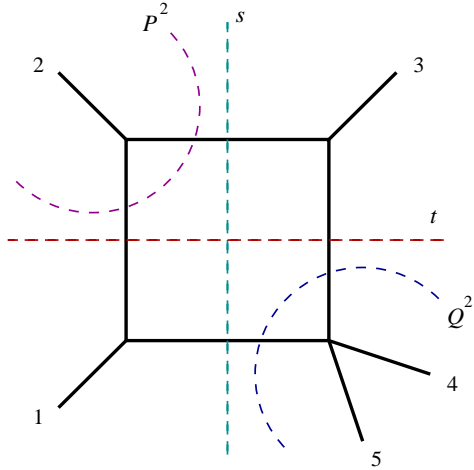}}
\end{center}
\caption{\it One of the box functions appearing in the expression of the one-loop
amplitude $\cM^{1-{\rm loop}}(1^- 2^-3^+ 4^+ 5^+) $.
}
\label{box5}
\end{figure}

Performing integrations in  \eqref{5p} using the result \eqref{tagl},  we see that
the various terms appearing in \eqref{5p} give  $P_{123}^2$-channel
dispersion integrals of cut-boxes.
A similar procedure will be followed for all the remaining MHV diagrams.
One then sums over all MHV diagrams, collecting  contributions to the same box function
arising from the different diagrams.

As an example,  let us  focus on the reconstruction of
the box integral  in Figure \ref{box5}.
One needs to sum  the three contributions
from the function $R(31)$ in the first three terms of \eqref{5p},
and the contribution from the function $R(13)$ in the second
term of \eqref{5p}. These will appear with a coefficient
\beq
\frac{\lan 12 \ran^8}{\lan 12 \ran^2 \lan 23 \ran^2} \frac{[45] \ \
s_{12} s_{23}}{\lan 14 \ran
\lan 15 \ran
\lan 34 \ran \lan 35 \ran \lan 45 \ran} \ ,
\eeq
which is precisely what expected from  the result derived in
\cite{Bern:1998sv}.%
\footnote{In order to match our result to that in \cite{Bern:1998sv},
one should remember the relation between  the box functions
$F^{123(45)} =  s_{12}s_{23} \, \cI^{123(45)}$. }

One should then consider the contributions to  this box function
from the MHV diagrams in the  null-cuts.
In appendix \ref{appendixB} we argue, following \cite{bst}, that
specific choices of $\eta$ allow to completely discard such diagrams.
Using this procedure, we have checked that our result for the
five-point amplitude  $\cM^{1-{\rm loop}} (1^- 2^- 3^+ 4^+ 5^+)$
precisely agrees with that of \cite{Bern:1998sv}.

\section{General procedure for  ${n}$-point amplitudes and \\ conclusions}

Finally, we outline a step-by-step procedure
which can  be applied to deal with MHV diagrams corresponding
to MHV amplitudes with an arbitrary number of particles.

The building blocks of the new set of diagrammatic rules are
gravity MHV amplitudes, appropriately continued to off-shell vertices.
MHV amplitudes of gravitons are not holomorphic in the spinor variables,
hence in section 2  we have supplied  a prescription for associating spinors
-- specifically  the  anti-holomorphic spinors --
to the  loop momenta. This prescription is defined by certain
shifts \eqref{shift}, which we rewrite here for convenience:
\beqa
\hat{\tilde{l}}_1 \ = \ \tilde{l}_1-z \, \frac{\lan l_2 \eta \ran}{\lan l_1 l_2 \ran} \,
\tilde{\eta} \ , \nonumber \\
\hat{\tilde{l}}_2 \ = \ \tilde{l}_2-z \, \frac{\lan l_1 \eta \ran}{\lan l_1 l_2 \ran} \,
\tilde{\eta}
\ .
\label{shiftagg}
\eeqa
These shifts are engineered in such a way to preserve momentum conservation
at the MHV vertices, and therefore give us the possibility of choosing
as MHV vertex any of the equivalent forms of the tree-level
amplitudes. The calculation of a one-loop MHV amplitude with an arbitrary
number of external legs is a straightforward generalisation of 
the four- and five-graviton cases discussed earlier, and 
proceeds along the following steps:
\begin{enumerate}
\item
Write the expressions for all relevant MHV diagrams, using
tree-level MHV vertices with shifted loop momenta
given by \eqref{shift}. The expression for these vertices
can be obtained by e.g.~applying  the appropriate KLT relations.
When required, sum over the particles of the supermultiplet which can
run in the loop.

\item
If a diagram has a null  two-particle cut, one applies  momentum conservation
of the three-point amplitude in order to cancel the presence of unphysical double poles.
Our calculations (and similar ones in Yang-Mills \cite{bst,bbst1,bbst2})
show that these diagrams give  a zero contribution upon choosing the gauge in
an appropriate way; thus they can be discarded (see appendix \ref{appendixB} for
a discussion of this point).

\item
Use  momentum conservation (with the shifts in place) in order to eliminate
any dependence on shifted momenta.
Once the integral is expressed entirely in terms of unshifted quantities,
one can apply any  reduction technique in order to produce
an expansion in terms of boxes and, possibly,
bubbles and triangles (which in $\cN=8$  should cancel \cite{zero}).

\item
Perform  the dispersive integrations as in section \ref{calc},
sum contributions from all MHV diagrams which can be built from MHV vertices,
and finally reconstruct each  box  as a  sum of four dispersion integrals -- in
its $s$-, $t$-, $P^2$- and $Q^2$-channels, using  \eqref{sumimp}.

\end{enumerate}

Clearly, it would be desirable to derive our prescription to continue off shell the loop momenta
from first principles. In particular,  it would be very interesting to find a derivation of 
the MHV diagram method in gravity similar to that of \cite{Mansfield,Gorsky:2005sf},  by performing an appropriate change of variables 
which would map the lightcone gravity action of  \cite{Scherk:1974zm} into an infinite sum of vertices, 
local in lightcone time, 
each with  the MHV helicity structure.
It would also be interesting if the MHV diagram description for gravity could be related, at least heuristically, 
to twistor string formulations of supergravity theories, 
such as those considered in  \cite{Abou-Zeid:2006wu}.  
We also notice that using the same shifts as in \eqref{shiftagg},  one should be able to 
perform a calculation of one-loop MHV amplitudes of gravitons in theories with less supersymmetry. 
For pure gravity,  rational terms in the amplitudes are not a priori correctly reproduced by the MHV diagram method, 
similarly to non-supersymmetric Yang-Mills.
For instance, pure gravity has an infinite sequence of all-plus graviton 
amplitudes  which are  finite and rational. As for the the all-plus gluon amplitudes in non-supersymmetric 
Yang-Mills theory,  it is conceivable that the all-plus graviton amplitudes 
arise in the MHV diagram method through violations of the $S$-matrix equivalence theorem 
in dimensional  regularisation \cite{Ettle:2007qc},  or from four-dimensional helicity-violating 
counterterms as in \cite{bstz}. 


                \section*{Acknowledgements}

It is a pleasure to thank  James Bedford, 
Paul Heslop, Costas Zoubos and especially Andi Brandhuber and Bill Spence 
for discussions.
The work of GT is supported by an EPSRC Advanced Fellowship EP/C544242/1
and by an EPSRC Standard Research Grant EP/C544250/1.

\newpage

\startappendix

\section{Comments on diagrams with null cuts}
\label{appendixB}

In this appendix we  would like to reconsider the contributions
to the MHV amplitudes arising from MHV diagrams with  a null two-particle cut.

\begin{figure}[ht]
\begin{center}
\scalebox{0.7}{\includegraphics{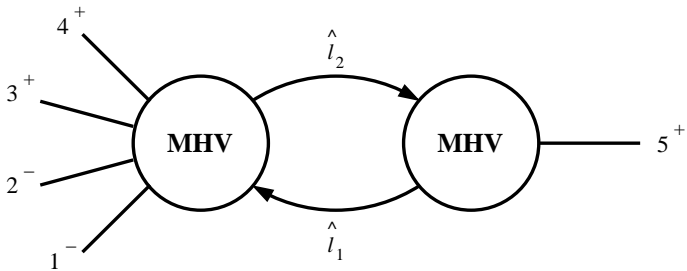}}
\end{center}
\caption{\it MHV diagram with null two-particle cut contributing
to the five-point graviton MHV amplitude at one loop.
}
\label{quattroeuno}
\end{figure}

An example is the MHV diagram in Figure \ref{quattroeuno}, contributing to
the five-point MHV amplitude discussed in section \ref{5ptampl}.
The expression for this diagrams is
\beqa
\label{pmd}
\cM^{1-{\rm loop}} \ = \ \int\!  d\m_{k_5} \
\cM( -\hat{l}_1^+1^- 2^- 3^+\hat{l}_2^+4^+) \ \cM(\hat{l}_1^- -\hat{l}_2^- 5^+)
\  .
\eeqa

Using  KLT relations for six-  \eqref{6} and for three-graviton amplitudes
\eqref{3}, we can write  \eqref{pmd} as a  sum of
two terms plus permutations of the particles $\cP(123)$.
Similarly to  section \ref{subsecnull},
momentum conservation $k_5-\hat{l}_2+\hat{l}_1 = 0 $ allows to prove
easily  the cancellation  of unphysical double poles
appearing because of the presence of a three-point graviton vertex.
Furthermore, all the dependence on hatted quantities can
be eliminated using momentum conservation in the form
\beq
\lan l_1 l_2 \ran [\hat{l}_2 i] \ = \ \lan l_1 5 \ran [5 i] \ , \qquad
\lan l_2 l_1 \ran [\hat{l}_1 j] \ = \ - \lan l_2 5 \ran [5 j] \ .
\eeq
Following this procedure,  the starting expression \eqref{pmd}
is decomposed into a sum of terms,
on which one easily applies  PV reduction techniques.
Similarly to the four-point case, one can see that only box functions in null  cuts
are produced.

The remark we would like to make now is that such terms  actually  vanish
with appropriate choices of the null reference vector $\eta$,
as observed in the Yang-Mills case in \cite{bst}. The same choice of $\eta$ has been used in
\cite{bbst1,quig,bbst2} in deriving gluon amplitudes in Yang-Mills theory, and recently
 in \cite{Nigel1,Nigel2,Nigel3} in deriving  one-loop $\phi$-MHV amplitudes,
 i.e.~amplitudes with gluons in an MHV helicity configuration and a complex scalar
 $\phi$ coupled to the gluons via the interaction $\phi\, {\rm Tr} F_{\mu \nu} F^{\mu \nu}$.

In \cite{bst},   it was found how a generic two-mass easy box function is reconstructed
by summing over four dispersion integrals,  as in  \eqref{sumimp}.
These dispersion integrals are performed in the four  channels
$s$, $t$, $P^2$ and $Q^2$ of the box function. As explained in that paper, the evaluation of
these integrals is greatly facilitated by choosing the reference vector $\eta$ appearing in
\eqref{off} to be   one of the two massless momenta, $p$ and $q$, of the box function
(see Figure   \ref{boxgen} for the labeling of the momenta in
a generic two-mass easy box).
By performing this choice, one finds that  the contribution of a single dispersion integral
of a cut-box  in a generic cut $s_{\rm cut}$ is proportional,
to all orders in the dimensional regularisation parameter $\epsilon$, to  \cite{ftt}
\beq
\label{plsc}
 -\frac{c_{\Gamma}}{\epsilon^2}  (-s_{\rm cut})^{- \epsilon} \,
{}_2F_{1}(1, - \epsilon, 1- \epsilon, a s_{\rm cut}) \ ,
\eeq
where $c_{\Gamma}$ is defined in \eqref{cgamma} and
$a$  is defined in \eqref{adef}.
In the four-point box function,  one obviously has  $P^2=Q^2=0$.
Using \eqref{plsc}, it is then immediate to see that
the dispersion integrals in these two channels vanish because of the presence
of the factor $(-s_{\rm cut})^{- \epsilon}$.
Therefore, when summing over all the
possible MHV diagrams,  it is in fact enough  to consider only
the MHV diagrams with non-vanishing cuts.

Finally, we notice that for arbitrary choices of $\eta$,
this would no longer be true; the MHV diagrams in null channels
would be important to restore $\eta$-independence in the final expressions
of one-loop amplitudes.

As a side remark, it instructive to apply the above comments
to rederive with MHV diagrams, almost instantly,  the expression
to all orders in the dimensional regularisation parameter, $\epsilon$, of the
one-loop  four-gluon amplitude in $\cN =4$ super Yang-Mills.
In this case, the result comes from summing two dispersion integrals, namely those in the
$s=(k_1+k_2)^2$ and in the $t=(k_2+k_3)^2$ channels;
indeed,  the specific choices of $\eta$ mentioned above allow us to
discard the MHV diagrams with null  two-particle cut.
In the four-particle case, the expression for
$a$ in \eqref{adef} simplifies to $a|_{P^2 = Q^2 = 0} = 1/s + 1/t$.
One then quickly  obtains, to all orders in $\epsilon$ \cite{ftt}, 
\beq
\label{4ptymeps}
\cA^{1-{\rm loop}} \ = \ 2 \cA^{\rm tree} \, \frac{c_{\Gamma}}{\epsilon^2} \left[
(-s)^{- \epsilon}
{}_2F_{1}\left( 1, - \epsilon, 1- \epsilon, 1+\frac{s}{t} \right) +
(-t)^{- \epsilon}
{}_2F_{1} \left( 1, - \epsilon, 1- \epsilon, 1+\frac{t}{s} \right) \right] \ .
\eeq
\eqref{4ptymeps} agrees  with the known result \cite{Green:1982sw}.

\section{Reduction technique of the $R$-functions}
\label{appa}

In dealing  with expressions of  gravity amplitudes derived using the MHV diagram method,
one often encounters products of  ``$R$-functions", where
\beq
R(ij) \ = \ \frac{\lan i l_2 \ran \lan j l_1 \ran}{\lan i l_1 \ran \lan j l_2 \ran} \ .
\eeq
The appearance of products of these functions is related to the structure
of tree-level gravity amplitudes, which can be expressed, using KLT relations,
as sums of products of two Yang-Mills amplitudes.
Here we would like to discuss how to reduce products of $R$-functions to
sums of $R$-functions and bubbles.

To begin with, we observe some useful properties of these functions:
\beqa
R(ab)R(bc) \ = \ R(ac)  \ \Rightarrow \ R(ab)R(ba) \ = \ 1 \ , \\ \cr
R(ab)R(cd) \ = \ R(ad)R(cb) \ \Rightarrow \ R(ab)R(da) \ = \ R(db) \ .
\eeqa
Let us now consider a generic  product $R(ij)R(hk)$  with $i \neq j \neq h \neq k$,
\beq
R(ij)R(hk) \ = \ \frac{\lan i l_2 \ran \lan j l_1 \ran}{\lan i l_1 \ran \lan j l_2 \ran}
\frac{\lan h l_2 \ran \lan k l_1 \ran}{\lan h l_1 \ran \lan k l_2 \ran}
\ .
\eeq
Using Schouten's identity in the form
\beq
\frac{\lan a l \ran}{\lan bl \ran \lan cl \ran} \ = \
\frac{\lan ac \ran}{\lan bc \ran} \frac{1}{\lan cl \ran} +
\frac{\lan ba \ran}{\lan bc \ran} \frac{1}{\lan bl \ran}
\ ,
\eeq
one can  separate contributions from different poles.
Applying this to the two ratios
$
\lan k l_1 \ran / (\lan i l_1 \ran \lan h l_1 \ran)
$
and
$
\lan h l_2 \ran / (\lan j l_2 \ran \lan k l_2 \ran)$,
we get
\beq \label{RR}
R(ij)R(hk) \ = \ \frac{\lan ik \ran \lan jh \ran}{\lan ih \ran \lan jk \ran}R(ij)+
\frac{\lan hk \ran}{\lan ih \ran \lan jk \ran} \left[ \lan kh \ran \frac{\mathcal{K} (ij)}{\mathcal{K}(kh)}+
\lan hj \ran \frac{\mathcal{K}(ij)}{\mathcal{K}(jh)}+ \lan ik \ran \frac{\mathcal{K}(ij)}{\mathcal{K}(ki)} \right]
\ ,
\eeq
where we have defined
\beq
\mathcal{K}(ij) \ := \ \lan i l_2 \ran \lan j l_1 \ran \ .
\eeq
Notice that  $R(ij)$ can be expressed in terms of $\mathcal{K}_{ij}$ as
\beq
R(ij) \ = \ \frac{\mathcal{K}(ij)}{\mathcal{K}(ji)}
\eeq
We can use again the same decomposition on a generic term
\beq
\frac{\mathcal{K}(ij)}{\mathcal{K}(hk)} \ = \
\frac{\lan i l_2 \ran \lan j l_1 \ran}{\lan h l_2 \ran \lan k l_1 \ran}
\frac{\lan h l_1 \ran}{\lan h l_1 \ran} \frac{\lan k l_2 \ran}{\lan k l_2 \ran}
\ ,
\eeq
to  get
\beq
\frac{\mathcal{K}(ij)}{\mathcal{K}(hk)} \ = \ \frac{\lan kj \ran \lan hi \ran}{\lan kh \ran \lan hk \ran}R(kh)+
\frac{1}{\lan kh \ran \lan hk \ran} \left[\lan jh \ran \lan ik \ran + \lan jh \ran \lan hi \ran
\frac{\lan k l_2 \ran}{\lan h l_2 \ran}+ \lan kj \ran \lan ik \ran \frac{\lan h l_1 \ran}{\lan k l_1 \ran}
\right]
\ .
\eeq
By substituting this expression into  \eqref{RR}, we see that we are
left with a bubble plus the sum of $R$-functions.
Using the Schouten identity,  we arrive at the final result
\beq
R(ij)R(hk) \ = \ -1+ \frac{\lan hk \ran \lan ij \ran}{\lan ih \ran \lan jk \ran} \left[R(hj)+R(ik) \right]+
\frac{\lan ik \ran \lan jh \ran}{\lan ih \ran \lan jk \ran} \left[ R(ij)+R(hk) \right]
\ .
\eeq
This formula allows us to perform immediately PV reductions of $R$-functions.
Further reducing the $R$-functions as usual \eqref{erre2}, we are then  left with bubbles, triangles and boxes.

\section{KLT relations}
For completeness, in this appendix
we include the field theory limit expressions of the KLT relations
\cite{klt} for the case of four-, five- and six-point amplitudes.
These are,
\beqa
\label{3}
\cM (1,2,3) & = &
-i\cA (1,2,3)\, \cA (1,2,3)
\ ,
\\  \cr
\cM (1,2,3,4) & = &
-is_{12}\ \cA (1,2,3,4)\cA(1,2,4,3) \ ,
\label{4}
\eeqa
\beqa
\label{5}
\cM (1,2,3,4,5)  & = &
is_{12}s_{34}\ \cA (1,2,3,4,5)\cA(2,1,4,3,5)
\nonumber \\
&+&i s_{13}s_{24}\ \cA(1,3,2,4,5)\cA(3,1,4,2,5) \ ,
\\ \cr
\label{6}
\cM (1,2,3,4,5,6)& = &
-is_{12}s_{45}\ \cA(1,2,3,4,5,6)
\big[ s_{35}\cA(2,1,5,3,4,6)
\nonumber \\
 & + & (s_{34}+s_{35})\ \cA (2,1,5,4,3,6)\big]
\\ \nonumber &+ & \cP (2,3,4) \ .
\eeqa
In these formulae, $\cM$ ($\cA$) denotes
a tree-level gravity (Yang-Mills, colour-ordered)
amplitude, $s_{ij} : = (k_i + k_j)^2$,
and $\cP (2,3,4)$ stands for permutations of $(2,3,4)$.
The form of KLT relations for a generic number of particles can be found in
 \cite{Bern:1998sv}.



\newpage
\begin{thebibliography}{99}

 \bibitem{witten} E.~Witten,
{\it Perturbative gauge theory as a string theory in twistor space}, Commun.\ Math.\ Phys.\
{\bf 252}, 189 (2004),  {\tt hep-th/0312171}.


\bibitem{Cachazo:2005ga}
F.~Cachazo and P.~Svr\v{c}ek, {\it Lectures on twistor strings and perturbative Yang-Mills theory},
PoS {\bf RTN2005}, 004 (2005),  {\tt hep-th/0504194}.

\bibitem{csw} F.~Cachazo, P.~Svr\v{c}ek and E.~Witten,
{\it MHV vertices and tree amplitudes in gauge theory}, JHEP {\bf 0409} (2004) 006, {\tt
hep-th/0403047}.




\bibitem{v1}
  G.~Georgiou and V.~V.~Khoze,
  {\it Tree amplitudes in gauge theory as scalar MHV diagrams,}
  JHEP {\bf 0405} (2004) 070
  {\tt hep-th/0404072}.

\bibitem{z1}
  J.~B.~Wu and C.~J.~Zhu,
 {\it MHV vertices and fermionic scattering amplitudes in gauge theory with
  quarks and gluinos,}
  JHEP {\bf 0409} (2004) 063
  {\tt hep-th/0406146}.

\bibitem{z2}
  J.~B.~Wu and C.~J.~Zhu,
  {\it MHV vertices and scattering amplitudes in gauge theory,}
  JHEP {\bf 0407} (2004) 032,
  {\tt hep-th/0406085}.


\bibitem{v2}
  G.~Georgiou, E.~W.~N.~Glover and V.~V.~Khoze,
  {\it Non-MHV tree amplitudes in gauge theory,}
  JHEP {\bf 0407} (2004) 048,
  {\tt hep-th/0407027}.



\bibitem{Mansfield}
 P.~Mansfield,
 {\it The Lagrangian origin of MHV rules,}
 JHEP {\bf 0603} (2006) 037,
{\tt hep-th/0511264}.


\bibitem{Gorsky:2005sf}
  A.~Gorsky and A.~Rosly,
  {\it From Yang-Mills Lagrangian to MHV diagrams,}
  JHEP {\bf 0601} (2006) 101,
  {\tt hep-th/0510111}.


\bibitem{Ettle:2006bw}
  J.~H.~Ettle and T.~R.~Morris,
  {\it Structure of the MHV-rules Lagrangian,}
  JHEP {\bf 0608}, 003 (2006),
  {\tt hep-th/0605121}.

\bibitem{bms}
  R.~Boels, L.~Mason and D.~Skinner,
 {\it From twistor actions to MHV diagrams,}
  Phys.\ Lett.\  B {\bf 648} (2007) 90, 
  {\tt hep-th/0702035}.


\bibitem{bcf}
R.~Britto, F.~Cachazo and B.~Feng,
{\it New recursion relations for tree amplitudes of gluons,}
Nucl.\ Phys.\ B {\bf 715}, 499 (2005)
{\tt hep-th/0412308}.

 \bibitem{bcfw}
  R.~Britto, F.~Cachazo, B.~Feng and E.~Witten,
  {\it Direct proof of tree-level recursion relation in Yang-Mills theory,}
  Phys.\ Rev.\ Lett.\  {\bf 94} (2005) 181602,
  {\tt hep-th/0501052}.



\bibitem{bbst3}
  J.~Bedford, A.~Brandhuber, B.~Spence and G.~Travaglini,
  {\it A recursion relation for gravity amplitudes,}
  Nucl.\ Phys.\ B {\bf 721}, 98 (2005),
  {\tt hep-th/0502146}.

\bibitem{cs}
  F.~Cachazo and P.~Svr\v{c}ek,
 {\it Tree level recursion relations in general relativity,}
  {\tt hep-th/0502160}.


\bibitem{bmst2}
  A.~Brandhuber, S.~McNamara, B.~Spence and G.~Travaglini,
  {\it Recursion relations for one-loop gravity amplitudes,}
  JHEP {\bf 0703} (2007) 029, 
 {\tt hep-th/0701187}.



\bibitem{Benincasa:2007qj}
  P.~Benincasa, C.~Boucher-Veronneau and F.~Cachazo,
  {\it Taming tree amplitudes in general relativity,}
  {\tt hep-th/0702032}.

\bibitem{Bjerrum-Bohr:2005xx}
  N.~E.~J.~Bjerrum-Bohr, D.~C.~Dunbar and H.~Ita,
  {\it Six-point one-loop N = 8 supergravity NMHV amplitudes and their IR
  behaviour,}
  Phys.\ Lett.\  B {\bf 621} (2005) 183,
  {\tt hep-th/0503102}.




\bibitem{zero}
  N.~E.~J.~Bjerrum-Bohr, D.~C.~Dunbar, H.~Ita, W.~B.~Perkins and K.~Risager,
 {\it The no-triangle hypothesis for N = 8 supergravity,}
  JHEP {\bf 0612} (2006) 072,
  {\tt hep-th/0610043}.


\bibitem{zerouno}
  Z.~Bern, N.~E.~J.~Bjerrum-Bohr and D.~C.~Dunbar,
 {\it Inherited twistor-space structure of gravity loop amplitudes,}
  JHEP {\bf 0505} (2005) 056, 
  {\tt hep-th/0501137}.




\bibitem{Green:2006gt}
  M.~B.~Green, J.~G.~Russo and P.~Vanhove,
  {\it Non-renormalisation conditions in type II string theory and maximal supergravity},
  {\tt hep-th/0610299}.

\bibitem{Bern:2006kd}
  Z.~Bern, L.~J.~Dixon and R.~Roiban,
  {\it Is N = 8 supergravity ultraviolet finite?},
  {\tt hep-th/0611086}.

\bibitem{Green:2006yu}
  M.~B.~Green, J.~G.~Russo and P.~Vanhove,
  {\it Ultraviolet properties of maximal supergravity,}
  {\tt hep-th/0611273}.


\bibitem{Bern:2007hh}
  Z.~Bern, J.~J.~Carrasco, L.~J.~Dixon, H.~Johansson, D.~A.~Kosower and R.~Roiban,
 {\it Three-loop superfiniteness of N = 8 supergravity,}
  {\tt hep-th/0702112}.

\bibitem{chalmers}
  G.~Chalmers,
  {\it On the finiteness of N = 8 quantum supergravity,}
{\tt hep-th/0008162}.





\bibitem{bdipr}
  N.~E.~J.~Bjerrum-Bohr, D.~C.~Dunbar, H.~Ita, W.~B.~Perkins and K.~Risager,
  {\it MHV-vertices for gravity amplitudes,}
  JHEP {\bf 0601} (2006) 009,
  {\tt hep-th/0509016}.



\bibitem{grav1}
S.~Giombi, R.~Ricci, D.~Robles-Llana and D.~Trancanelli,
{\it A note on twistor gravity amplitudes},
JHEP {\bf 0407} (2004) 059,  {\tt hep-th/0405086}.




\bibitem{ris}
  K.~Risager,
  {\it A direct proof of the CSW rules,}
  JHEP {\bf 0512} (2005) 003,
  {\tt hep-th/0508206}.

\bibitem{bst}
A.~Brandhuber, B.~Spence and G.~Travaglini, {\it One-Loop Gauge Theory Amplitudes in N=4 super
Yang-Mills from MHV Vertices}, Nucl.\ Phys.\ B {\bf 706} (2005) 150, {\tt  hep-th/0407214}.


\bibitem{BT06}
A.~Brandhuber and G.~Travaglini, {\it Quantum {MHV} diagrams},
{{\tt hep-th/0609011}}.





\bibitem{bddk}
Z.~Bern, L.~J.~Dixon, D.~C.~Dunbar and D.~A.~Kosower, {\it Fusing gauge theory tree amplitudes
into loop amplitudes,} Nucl.\ Phys.\ B {\bf 435} (1995) 59, {\tt hep-ph/9409265}.


\bibitem{bddkcoll}
  Z.~Bern, L.~J.~Dixon, D.~C.~Dunbar and D.~A.~Kosower,
  {\it One loop n point gauge theory amplitudes, unitarity and collinear limits,}
  Nucl.\ Phys.\  B {\bf 425} (1994) 217, 
  {\tt hep-ph/9403226}.


\bibitem{ftt}
  A.~Brandhuber, B.~Spence and G.~Travaglini,
  {\it From trees to loops and back,}
  JHEP {\bf 0601} (2006) 142,
  {\tt hep-th/0510253}.



\bibitem{F1}
  R.~P.~Feynman,
  {\it Quantum Theory Of Gravitation,}
  Acta Phys.\ Polon.\  {\bf 24} (1963) 697.




\bibitem{F2}
  R.~P.~Feynman,
  {\it Closed Loop And Tree Diagrams,}
 in J.~R.~Klauder, {\it Magic Without Magic},
 San Francisco 1972, 355-375; in  Brown, L.~M.~(ed.):
 {\it Selected papers of Richard Feynman,} 867-887



\bibitem{quig}
C.~Quigley and M.~Rozali, {\it One-Loop MHV Amplitudes in Supersymmetric Gauge Theories},
 JHEP {\bf 0501} (2005) 053, 
  {\tt hep-th/0410278}.

\bibitem{bbst1}
J.~Bedford, A.~Brandhuber, B.~Spence and G.~Travaglini, {\it A Twistor Approach to One-Loop
Amplitudes in ${\cal N} \! = \! 1$ Supersymmetric Yang-Mills Theory}, Nucl.\ Phys.\ B {\bf 706}
(2005) 100, {\tt hep-th/0410280}.

\bibitem{bbst2}
J.~Bedford, A.~Brandhuber, B.~Spence and G.~Travaglini, {\it Non-supersymmetric loop amplitudes
and MHV vertices,} Nucl.\ Phys.\ B {\bf 712} (2005) 59, {\tt hep-th/0412108}.





\bibitem{Nigel1}
  S.~D.~Badger and E.~W.~N.~Glover,
  {\it One-loop helicity amplitudes for H $\to$  gluons: The all-minus
  configuration,}
  Nucl.\ Phys.\ Proc.\ Suppl.\  {\bf 160} (2006) 71, {\tt hep-ph/0607139}.

\bibitem{Nigel2}
  S.~D.~Badger, E.~W.~N.~Glover and K.~Risager,
  {\it One-loop phi-MHV amplitudes using the unitarity bootstrap,}
  {\tt 0704.3914 [hep-ph]}.


\bibitem{Nigel3}
  S.~D.~Badger, E.~W.~N.~Glover and K.~Risager,
  {\it Higgs amplitudes from twistor inspired methods,}
  {\tt 0705.0264 [hep-ph].}


\bibitem{LanceNigelValya}
  L.~J.~Dixon, E.~W.~N.~Glover and V.~V.~Khoze,
  {\it MHV rules for Higgs plus multi-gluon amplitudes,}
  JHEP {\bf 0412} (2004) 015,
  {\tt hep-th/0411092}.


\bibitem{Badger:2004ty}
  S.~D.~Badger, E.~W.~N.~Glover and V.~V.~Khoze,
 {\it MHV rules for Higgs plus multi-parton amplitudes,}
  JHEP {\bf 0503} (2005) 023, 
  {\tt hep-th/0412275}.



\bibitem{pureYM}
  A.~Brandhuber, B.~Spence and G.~Travaglini,
  {\it Amplitudes in pure Yang-Mills and MHV diagrams,}
JHEP {\bf 0702} (2002) 88,
  {\tt hep-th/0612007}.


\bibitem{Ettle:2007qc}
  J.~H.~Ettle, C.~H.~Fu, J.~P.~Fudger, P.~R.~W.~Mansfield and T.~R.~Morris,
  {\it S-Matrix Equivalence Theorem Evasion and Dimensional Regularisation with
  the Canonical MHV Lagrangian,}
  {\tt hep-th/0703286}.

\bibitem{bstz}
  A.~Brandhuber, B.~Spence, G.~Travaglini and K.~Zoubos,
  {\it One-loop MHV Rules and Pure Yang-Mills,}
  {\tt 0704.0245 [hep-th]}.



\bibitem{Green:1982sw}
  M.~B.~Green, J.~H.~Schwarz and L.~Brink,
 {\it N=4 Yang-Mills And N=8 Supergravity As Limits Of String Theories,}
  Nucl.\ Phys.\  B {\bf 198} (1982) 474.

\bibitem{dn}
D.~C.~Dunbar and P.~S.~Norridge,
{\it Calculation of graviton scattering amplitudes using string based methods},
Nucl.\ Phys.\ B {\bf 433}, 181 (1995),
{\tt hep-th/9408014}.

\bibitem{bk}
  Z.~Bern and D.~A.~Kosower,
  {\it The Computation of loop amplitudes in gauge theories,}
  Nucl.\ Phys.\  B {\bf 379} (1992) 451.



\bibitem{Bern:1998sv}
Z.~Bern, L.~J.~Dixon, M.~Perelstein and J.~S.~Rozowsky,
{\it Multi-leg one-loop gravity amplitudes from gauge theory,}
Nucl.\ Phys.\ B {\bf 546} (1999) 423,
{\tt hep-th/9811140}.



\bibitem{Bern:2002kj}
Z.~Bern,
{\it Perturbative quantum gravity and its relation to gauge theory,}
Living Rev.\ Rel.\  {\bf 5}, 5 (2002)
{\tt gr-qc/0206071}.



 \bibitem{david}
  D.~A.~Kosower,
  {\it Next-to-maximal helicity violating amplitudes in gauge theory,}
  Phys.\ Rev.\  D {\bf 71} (2005) 045007,
  {\tt hep-th/0406175}.



\bibitem{bgk}
  F.~A.~Berends, W.~T.~Giele and H.~Kuijf,
  {\it On relations between multi-gluon and multigraviton scattering,}
  Phys.\ Lett.\ B {\bf 211} (1988) 91.




\bibitem{mp}
  M.~L.~Mangano and S.~J.~Parke,
  {\it Multiparton amplitudes in gauge theories,}
  Phys.\ Rept.\  {\bf 200} (1991) 301,
  {\tt hep-th/0509223}.



\bibitem{gpvn}
  M.~T.~Grisaru, H.~N.~Pendleton and P.~van Nieuwenhuizen,
  {\it Supergravity And The S Matrix,}
  Phys.\ Rev.\ D {\bf 15} (1977) 996.






\bibitem{klt}
H.~Kawai, D.~C.~Lewellen and S.~H.~H.~Tye,
{\it A Relation Between Tree Amplitudes Of Closed And Open Strings,}
Nucl.\ Phys.\ B {\bf 269} (1986) 1.



\bibitem{Scherk:1974zm}
  J.~Scherk and J.~H.~Schwarz,
 {\it Gravitation In The Light - Cone Gauge,}
  Gen.\ Rel.\ Grav.\  {\bf 6} (1975) 537.


\bibitem{Abou-Zeid:2006wu}
  M.~Abou-Zeid, C.~M.~Hull and L.~J.~Mason,
  {\it Einstein supergravity and new twistor string theories,}
{\tt hep-th/0606272}.

\end{thebibliography}
\end{document}